\documentclass[conference]{IEEEtran}
\IEEEoverridecommandlockouts
\usepackage{amsmath,amssymb,amsfonts,amsthm,mathtools}
\usepackage{algorithmic}
\usepackage{graphicx}
\usepackage{textcomp}
\usepackage{xcolor}
\usepackage{amsmath,amsthm,enumitem,nicefrac,bbm,verbatim,amssymb,amsfonts,amscd, graphicx,hyperref,float,mathtools,bm,url,tikz,stackrel,subcaption,csquotes}
\newtheorem{Theorem*}{Theorem}

\newtheorem{Claim*}[Theorem]{Claim}

\newtheorem{CounterExample*}{$\overline{\hbox{\bf Example}}$}

\newtheorem{Example*}[Theorem]{Example}

\newtheorem{Intuition*}[Theorem]{Intuition}
\newtheorem{Joke*}[Theorem]{Joke}

\newtheorem{Lemma*}[Theorem]{Lemma}
\newtheorem{Open problem}[Theorem]{Open problem}

\newtheorem{Question*}[Theorem]{Question}

\usepackage{fullpage}

\newtheorem{theorem}{Theorem}

\newtheorem{definition}[theorem]{Definition}

\def \bSubexa    {\begin{subexa}}

\usepackage[colorinlistoftodos,textsize=scriptsize]{todonotes}

\newcommand{\ignore}[1]{}

\newcommand{\EE}{\mathbb{E}}

\newcommand{\NN}{\mathbb{N}}

\newcommand{\RR}{\mathbb{R}}

\newcommand{\PP}{\mathbb{P}}

\def \cC     {{\cal C}}

\def \cM     {{\cal M}}

\def \vct#1{{\overrightarrow{#1}}}

\def \vctZ  {{\vct Z}}

\def \sqnrm#1{{\left \Vert {#1}\right \rVert^2}}

\def \Paren#1{{\left({#1}\right)}}

\def \Brack#1{{\left[{#1}\right]}}

\def\ignore#1{}

\newcommand{\bi}{\begin{itemize}}
\newcommand{\ei}{\end{itemize}}
\def\orpro{\mathop{\mathchoice
   {\vee\kern-.49em\raise.7ex\hbox{$\cdot$}\kern.4em}
   {\vee\kern-.45em\raise.63ex\hbox{$\cdot$}\kern.2em}
   {\vee\kern-.4em\raise.3ex\hbox{$\cdot$}\kern.1em}
   {\vee\kern-.35em\raise2.2ex\hbox{$\cdot$}\kern.1em}}\limits}

\def\andpro{\mathop{\mathchoice
 {\wedge\kern-.46em\lower.69ex\hbox{$\cdot$}\kern.3em}
 {\wedge\kern-.46em\lower.58ex\hbox{$\cdot$}\kern.25em}
 {\wedge\kern-.38em\lower.5ex\hbox{$\cdot$}\kern.1em}
 {\wedge\kern-.3em\lower.5ex\hbox{$\cdot$}\kern.1em}}\limits}

\def\simge{\mathrel{%
   \rlap{\raise 0.511ex \hbox{$>$}}{\lower 0.511ex \hbox{$\sim$}}}}

\def\simle{\mathrel{
   \rlap{\raise 0.511ex \hbox{$<$}}{\lower 0.511ex \hbox{$\sim$}}}}

\providecommand{\email}[1]{\href{mailto:#1}{\nolinkurl{#1}\xspace}}

\newcommand{\eps}{\varepsilon}

\DeclareMathOperator{\sinc}{sinc}

\newcommand{\phase}{V}

\newcommand{\eqdef}{\stackrel{\text{def}}{=}}

\newcommand{\lebtwo}{L^2\left[0,1\right]}

\newcommand{\rmp}{J}

\newcommand{\circenc}{f_{\text{circ}}}

\newcommand{\rampenc}{f_{\text{ramp}}}
\newcommand{\rset}{S}
\newcommand{\ent}{H}
\newcommand{\leb}{\mu}
\newcommand{\dist}{D}

\newcommand{\sdim}{d_s}
\newcommand{\cdim}{d_c}
\newcommand{\fenc}{f}

\newcommand{\met}{\cM}
\newcommand{\darc}{\mu}
\newcommand{\anl}{g_a}
\newcommand{\syn}{g_s}
\newcommand{\angenc}{h}

\newcommand{\eramp}{E_{\text{r}}}
\newcommand{\ecirc}{E_{\text{c}}}

\usepackage[backend=biber, style=ieee, natbib=true, mincitenames=1, maxcitenames=2, giveninits=true, uniquename=init]{biblatex}
\AtBeginDocument{\toggletrue{blx@useprefix}}
\AtBeginBibliography{\togglefalse{blx@useprefix}}

\def\BibTeX{{\rm B\kern-.05em{\sc i\kern-.025em b}\kern-.08em
    T\kern-.1667em\lower.7ex\hbox{E}\kern-.125emX}}

\newcommand*{\EXTENDED}{}%
\allowdisplaybreaks[2]
\begin{document}

\title{Do Neural Networks Compress \\ Manifolds Optimally?
%	{\footnotesize \textsuperscript{*}Note: Sub-titles are not captured in Xplore and
%		should not be used}
    \thanks{The first two authors were supported by the US National Science Foundation
    under grants CCF-2008266 and CCF-1934985, by the US Army Research Office
    under grant W911NF-18-1-0426 and by a gift from Google.}}
\author{
    \IEEEauthorblockN{
    Sourbh Bhadane and Aaron B.~Wagner\\
    \IEEEauthorblockA{School of Electrical and Computer Engineering \\
    Cornell University \\
    Ithaca, NY 14853 USA \\
    snb62@cornell.edu, wagner@cornell.edu}}
%	\and
%    \IEEEauthorblockN{1\textsuperscript{st} Aaron B.~Wagner
%    \IEEEauthorblockA{School of Electrical and Computer Engineering} \\
%    Cornell University \\
%    Ithaca, NY 14853 USA \\
%    wagner@cornell.edu}
	\and
    \IEEEauthorblockN{
        Johannes Ball\'{e} \\
    \IEEEauthorblockA{Google Research \\
%    1600 Amphitheatre Pkwy \\
	 Mountain View, CA 94043 USA \\
     jballe@google.com}}}
%	\and
%	\IEEEauthorblockN{3\textsuperscript{rd} Given Name Surname}
%	\IEEEauthorblockA{\textit{dept. name of organization (of Aff.)} \\
%		\textit{name of organization (of Aff.)}\\
%		City, Country \\
%		email address or ORCID}
%	\and
%	\IEEEauthorblockN{4\textsuperscript{th} Given Name Surname}
%	\IEEEauthorblockA{\textit{dept. name of organization (of Aff.)} \\
%		\textit{name of organization (of Aff.)}\\
%		City, Country \\
%		email address or ORCID}
%	\and
%	\IEEEauthorblockN{5\textsuperscript{th} Given Name Surname}
%	\IEEEauthorblockA{\textit{dept. name of organization (of Aff.)} \\
%		\textit{name of organization (of Aff.)}\\
%		City, Country \\
%		email address or ORCID}
%	\and
%	\IEEEauthorblockN{6\textsuperscript{th} Given Name Surname}
%	\IEEEauthorblockA{\textit{dept. name of organization (of Aff.)} \\
%		\textit{name of organization (of Aff.)}\\
%		City, Country \\
%		email address or ORCID}
%}
%	\thanks{Identify applicable funding agency here. If none, delete this.}

\maketitle

\begin{abstract}
Artifical Neural-Network-based (ANN-based) lossy compressors have
recently obtained striking results on several sources. Their
success may be ascribed to an ability to identify the 
structure of low-dimensional manifolds in high-dimensional
ambient spaces. Indeed, prior work has shown that ANN-based
compressors can achieve the optimal entropy-distortion 
curve for some such sources. In contrast, we 
determine the optimal entropy-distortion tradeoffs for
two low-dimensional manifolds with circular structure
and show that state-of-the-art ANN-based compressors
fail to optimally compress them.
\end{abstract}


\section{Introduction}
Stochastically-trained
Artificial Neural-Network-based (ANN-based) lossy compressors are 
the state-of-the-art for various types of sources, most notably
images~\cite{BalleLS17,BalleMSHJ18,MentzerTTA20}. One particularly successful framework for using
ANNs for lossy compression, which is the focus of this paper,
 creates the compressed representation
by quantizing the latent variables in an autoencoder-like 
architecture~\cite{BalleCMSJAHT20}. One can view this as similar to the
transform-based coding approach that underpins JPEG and
other standards, except that the transforms are implemented
by multilayer perceptrons and can therefore be nonlinear.
These ANN-based compressors outperform linear-transform-based 
methods, even under a mean-squared-error (MSE)
distortion measure~\cite{BalleLS17}.  Since the latter is
provably near-optimal for stationary Gaussian sources
under an MSE distortion constraint, ANN-based compressors
are evidently able to exploit non-Gaussianity in sources.

Given that ensembles of images have long been suspected to 
live on low-dimensional manifolds in pixel-space, it is
natural to conjecture that ANN-based compressors are 
adept at compressing sources that exhibit low-dimensional
manifold structure in a high-dimensional ambient space.
Previous work~\cite{WaBa:Sawbridge:Long} considered a particular
random process (the ``sawbridge'') over $[0,1]$ exhibiting
this structure and
found that stochastically-trained ANNs indeed compress
the process optimally.

We consider the random process obtained
by applying a random cyclic shift to the function 
$t \mapsto t - 1/2$ over $[0,1]$. We call the resulting 
process the \emph{ramp}. We characterize the entropy-distortion
function for this process under an MSE distortion constraint.
Despite the considerable similarities between the ramp and
the sawbridge, we find that stochastically-trained ANN-based
compressors fail to compress the ramp optimally at high-SNR.
The difficulty stems from the fact that,
unlike the sawbridge, the set of ramp realizations forms
a closed loop in function space, which  creates a 
topological challenge related to the impossibility of
mapping a circle to a segment in a continuous and
invertible way \cite[Chap 2. Ex. 7]{Munkres00}. To illustrate 
this issue in arguably its simplest form, we begin
by considering the problem of compressing the unit circle in 
two-dimensions. We characterize the entropy-distortion function and 
again find that stochastically-trained ANNs are suboptimal
at high rates.

\section{Preliminaries}

%We study variable-rate, ANN-based compressors parametrized by the dimension of the source, $\sdim$, and the dimension of the code, $\cdim$. 
%\begin{definition}
%	A variable rate compressor $\comp(.)$ consists of an analysis-synthesis transform $(\anl,\syn)$ and quantizer $\quant$ where $\anl:\RR^{\sdim} \mapsto \RR^{\cdim}$ and $\syn:\RR^{\cdim} \mapsto \RR^{\sdim}$ and $\quant:\RR^{\cdim}\mapsto \RR^{\cdim}$. For an input $\ip$, the output of the analysis transform $\anl(\ip)$ is passed through the quantizer to get  the code $q(\anl(X))$ which is passed through the synthesis transform to get the reconstruction $\hat{X} = \syn(q(\anl(X)))$. We denote the composition $\syn \circ \quant \circ \anl$ by $\comp$.
%\end{definition}
%

For a source $X$ in a space $\met$, we define an encoder and its entropy and distortion. In this paper $\met$ will be either $\lebtwo$ or $\RR^2$.
In both cases, conditional expectations and norms are well-defined.
\begin{definition}
    An \emph{encoder} is a mapping $\fenc: \met \mapsto \NN$. Its \emph{entropy} and \emph{distortion} are given by 
	\begin{align*}
		H(\fenc) &= \sum\limits_{i \in \NN} -\Pr\left( \fenc(X) = i \right) \log \Paren{\Pr\Paren{\fenc(X)=i}} \\ 
		D(\fenc) &= \EE\Brack{\left \lVert X- \EE\Brack{X \mid \fenc(X) }\right \rVert^2},
	\end{align*}
respectively.
\end{definition} 

Note in particular that we consider mean squared error as the
distortion measure.
We shall characterize optimal compression performance via the 
\emph{entropy-distortion function.}

\begin{definition}
The entropy-distortion function of $X$ is 
\begin{align*}
	E(D) = &\inf\limits_{\fenc} H(\fenc) \\
	\quad &\textrm{s.t. } D(\fenc) \leq D.
\end{align*} 
\end{definition} 

We consider the entropy-distortion function instead of the more-conventional
rate-distortion function~\cite[Theorem 10.2.1]{Cover:IT2} because ANN-based compressors 
optimize entropy, which is known to be a lower bound to the expected codeword
length under optimal, one-shot prefix-free encoding~\cite[Theorem 5.4.1]{Cover:IT2}

%The reason for considering the entropy-distortion function instead 
%of the more-conventional rate-distortion function~\cite{Cover:IT2} is 
%two-fold. First, the ANNs-based compressors that we consider
%optimize the entropy of the coded representation for a given
%mean-squared error. Thus the entropy-distortion function 
%describes the relevant fundamental limit. Secondly, the 
%rate-distortion describes the fundamental limit when the
%compression is amortized over many independent realizations
%of the source, whereas our focus here is on one-shot
%compression. It is well-known that the entropy of the
%encoder's output is a lower bound to, and within one
%bit of, the expected codeword length under optimal
%prefix-free encoding~\cite{Cover:IT2}.

%We are often interested in extreme rate regimes in cases where we cannot completely characterize the entropy-distortion tradeoff for all possible distortions. In this paper, we are interested in the high-rate, low-distortion asymptotes of a source. 
%
%\begin{definition}
%The high-rate slope of the entropy-distortion curve of a source $X$ is denoted by $\hslope$ and defined as 
%	\begin{equation}
%		\hslope \eqdef \lim\limits_{D \rightarrow 0} \frac{E(D)}{\log\Paren{\frac{1}{D}}}. 
%	\end{equation}
%\end{definition}

\section{The Circle}\label{sec:circle}

The \emph{circle} is a 2-D source with a 1-D latent variable.
\begin{equation} \label{eq:circle_def}
    \theta \sim \text{Unif}\left[ 0, 2\pi \right), \vctZ = \left(\cos \theta, \sin \theta\right).
\end{equation}
%\begin{figure}%TODO: Update figure
%	\centering
%	\includegraphics[scale=0.3]{tmp/c_realz}
%	\caption{Circle: Sample realizations.}\label{fig:c_realz}
%\end{figure}

We first derive its optimal entropy-distortion tradeoff and then analyze the performance of ANN-based compressors.
\subsection{The Optimal Tradeoff}\label{subsec:circleRD}

%In this section, we consider the two-dimensional \emph{circle} source given by the following process, 
%
%\[ \theta \sim \text{Unif}\left[ 0, 2\pi \right), \left(X,Y\right) = \left(\cos \theta, \sin \theta\right).\]
\begin{theorem}\label{thm:circleED}
	For the circle, if $D \geq 1$, $\ecirc(D)=0$. If $0<D<1$, then
	\begin{align*}
        \ecirc(D) = \inf\limits_{\left\{ \theta_{i} \right\}_{i=1}^{\infty}} \quad -&\sum\limits_{i}^{\infty} \frac{\theta_i}{2\pi} \log\Paren{\frac{\theta_i}{2\pi}} \\ 
		\textrm{s.t. } \quad& \sum\limits_{i=1}^{\infty} \frac{\theta_i}{2\pi} \left( 1- \sinc^2\Paren{\frac{\theta_i}{2}} \right) \leq D, \\
		& \sum\limits_{i=1}^{\infty} \theta_i = 2\pi, \theta_i \geq 0 \text{ for all i},
	\end{align*}
    where 
    \begin{equation}
        \sinc(x) = \begin{cases}
            \frac{\sin(x)}{x} & \text{if $x \ne 0$} \\
            1 & \text{otherwise}.
        \end{cases}
    \end{equation}
\end{theorem}

\ifnum0\ifdefined\EXTENDED 1\fi \ifdefined\JOURNAL 1\fi >0
\begin{proof}
	Let $\fenc:\RR^2 \mapsto \NN$ be an encoder. For the circle source, an encoder $f$ can be alternatively represented as a function of the angle $\theta \in \left[0,2\pi \right)$ as $\angenc(\theta) = \fenc\Paren{\left[\cos \theta, \sin\theta \right]}$.
	Let $\cC$ denote the unit circle and define $C_i \eqdef \cC \cap \fenc^{-1}(i)$ and $\Theta_i \eqdef \left[0,2\pi \right) \cap \angenc^{-1}(i)$. $C_i$ is said to be contiguous if $\Theta_i$ is an interval or if it is of the form $\left[0,\theta_1 \right] \cup \left[ 2\pi - \theta_2, 2\pi \right)$. Let $\darc(C_i)$ be the Lebesgue of $C_i$. The entropy and distortion are given by
	\begin{align*}
		H(\fenc) &= \sum\limits_{i=1}^{\infty} -\frac{\darc(C_i)}{2\pi} \log\Paren{\frac{\darc(C_i)}{2\pi}}, \\
		D(\fenc) &= \sum\limits_{i=1}^{\infty} \frac{\darc(C_i)}{2\pi} \EE\Brack{\left \lVert \vctZ - \EE\Brack{\vctZ \mid \vctZ \in C_i}\right \rVert^2 \mid \vctZ \in C_i }
	\end{align*} 
respectively. 
    
We now prove that 
    \begin{equation}
        \label{eq:circ:target}
	\sqnrm{\EE\Brack{\vctZ \mid \vctZ \in C_i }} \le 
        \sinc^2\Paren{\frac{\darc(C_i)}{2}}.
    \end{equation}
To this end, note that we can write
\ifdefined \JOURNAL 
\begin{align*}   \EE\Brack{\left \lVert \vctZ - \EE\Brack{\vctZ \mid \vctZ \in C_i}\right \rVert^2 \mid \vctZ \in C_i } &= \EE\Brack{\left \lVert \vctZ \right \rVert^2 \mid \vctZ \in C_i} - \left \Vert \EE\Brack{\vctZ \mid \vctZ \in C_i} \right \rVert^2 \\
	&= 1 - \left \Vert \EE\Brack{\vctZ \mid \vctZ \in C_i} \right \rVert^2.
\end{align*}
\else
\begin{align*}   
	&\EE\Brack{\left \lVert \vctZ - \EE\Brack{\vctZ \mid \vctZ \in C_i}\right \rVert^2 \mid \vctZ \in C_i } \\
	&= \EE\Brack{\left \lVert \vctZ \right \rVert^2 \mid \vctZ \in C_i} - \left \Vert \EE\Brack{\vctZ \mid \vctZ \in C_i} \right \rVert^2 \\
	&= 1 - \left \Vert \EE\Brack{\vctZ \mid \vctZ \in C_i} \right \rVert^2.
\end{align*}
\fi 
First suppose that $C_i$ is contiguous with
corresponding $\Theta = \left[ \theta_l, \theta_r \right]$ 
where $\theta_l < \theta_r$.  Then we have
\ifdefined \JOURNAL 
\begin{align}
    \nonumber
	\sqnrm{\EE\Brack{\vctZ \mid \vctZ \in C }} 
	&= \Paren{\EE\Brack{\cos \theta \mid \theta \in \left[ \theta_l ,\theta_r \right]}}^2 + \Paren{\EE\Brack{\sin \theta \mid \theta \in \left[ \theta_l ,\theta_r \right]}}^2 \\
    \nonumber
	&= \frac{1}{\Paren{\theta_r - \theta_l}^2}\Paren{\Paren{\sin \theta_r - \sin \theta_l}^2 + \Paren{\cos \theta_l - \cos \theta_r}^2} \\
    \nonumber
	&= \frac{1}{\Paren{\theta_r - \theta_l}^2} \Paren{2-2\cos \Paren{\theta_r - \theta_l}}\\
    \nonumber
	&=  \frac{4}{\Paren{\theta_r - \theta_l}^2} \sin^2\Paren{\frac{\theta_r - \theta_l}{2}}  \\
    \nonumber
    & = \sinc^2\Paren{\frac{\theta_r - \theta_l}{2}} \\
    \label{eq:circ:contiguous}
    & = \sinc^2\Paren{\frac{\darc(C_i)}{2}}.
\end{align}
\else 
\begin{align}
    \nonumber
&\sqnrm{\EE\Brack{\vctZ \mid \vctZ \in C }} \\
&= \Paren{\EE\Brack{\cos \theta \mid \theta \in \left[ \theta_l ,\theta_r \right]}}^2 + \Paren{\EE\Brack{\sin \theta \mid \theta \in \left[ \theta_l ,\theta_r \right]}}^2 \\
\nonumber
&= \frac{1}{\Paren{\theta_r - \theta_l}^2}\Paren{\Paren{\sin \theta_r - \sin \theta_l}^2 + \Paren{\cos \theta_l - \cos \theta_r}^2} \\
\nonumber
&= \frac{1}{\Paren{\theta_r - \theta_l}^2} \Paren{2-2\cos \Paren{\theta_r - \theta_l}}\\
\nonumber
&=  \frac{4}{\Paren{\theta_r - \theta_l}^2} \sin^2\Paren{\frac{\theta_r - \theta_l}{2}}  \\
\nonumber
& = \sinc^2\Paren{\frac{\theta_r - \theta_l}{2}} \\
\label{eq:circ:contiguous}
& = \sinc^2\Paren{\frac{\darc(C_i)}{2}}.
\end{align}
\fi 
More generally, suppose $C_i$ is a finite, disjoint union of closed intervals. If
$C_i$ consists of a single interval then it is contiguous. Otherwise, let $C_l$ denote
one of the intervals comprising $C_i$ and let $C_r$  denote the remainder. We have
\ifdefined \JOURNAL 
\begin{equation}
\EE\Brack{\vctZ \mid \vctZ \in C_i} = \frac{\darc(C_l)}{\darc(C_l) + \darc(C_r)} \EE\Brack{\vctZ \mid \vctZ \in C_l} + \frac{\darc(C_r)}{\darc(C_l)+ \darc(C_r)} \EE\Brack{\vctZ \mid \vctZ \in C_r}.	
\end{equation}
\else 
\begin{align*}
	\EE\Brack{\vctZ \mid \vctZ \in C_i} &= \frac{\darc(C_l)}{\darc(C_l) + \darc(C_r)} \EE\Brack{\vctZ \mid \vctZ \in C_l} \\
	&\hspace{5mm}+ \frac{\darc(C_r)}{\darc(C_l)+ \darc(C_r)} \EE\Brack{\vctZ \mid \vctZ \in C_r}.	
\end{align*}
\fi
and therefore,
\ifdefined \JOURNAL 
\begin{align*}
\left \Vert \EE\Brack{\vctZ \mid \vctZ \in C_i} \right \rVert^2 &= \Paren{\frac{\darc(C_l)}{\darc(C_l) + \darc(C_r)}}^2 \sqnrm{\EE\Brack{\vctZ \mid \vctZ \in C_l}} + \Paren{\frac{\darc(C_r)}{\darc(C_l) +\darc(C_r)}}^2 \sqnrm{\EE\Brack{\vctZ \mid \vctZ \in C_r}} \\
&\hspace{10mm}  + 2 \Paren{\frac{\darc(C_l)}{\darc(C_l) + \darc(C_r)}}^2 \Paren{\frac{\darc(C_r)}{\darc(C_l) +\darc(C_r)}}^2
    \EE\Brack{\vctZ \mid \vctZ \in C_l}^\top \EE\Brack{\vctZ \mid \vctZ \in C_r}.
\end{align*}
\else 
\begin{align*}
	&\left \Vert \EE\Brack{\vctZ \mid \vctZ \in C_i} \right \rVert^2 \\
	&= \Paren{\frac{\darc(C_l)}{\darc(C_l) + \darc(C_r)}}^2 \sqnrm{\EE\Brack{\vctZ \mid \vctZ \in C_l}} \\
	&\hspace{5mm}+ \Paren{\frac{\darc(C_r)}{\darc(C_l) +\darc(C_r)}}^2 \sqnrm{\EE\Brack{\vctZ \mid \vctZ \in C_r}} \\
	&\hspace{5mm} + 2 \Paren{\frac{\darc(C_l)}{\darc(C_l) + \darc(C_r)}}^2 \Paren{\frac{\darc(C_r)}{\darc(C_l) +\darc(C_r)}}^2 \times \\
	& \hspace{10mm} \EE\Brack{\vctZ \mid \vctZ \in C_l}^\top \EE\Brack{\vctZ \mid \vctZ \in C_r}.
\end{align*}
\fi 
The right-hand side is evidently maximized by rotating $C_l$ along the circle until it abuts
$C_r$. Repeating this process results in a contiguous $C_i$, for which (\ref{eq:circ:contiguous})
holds. 

Finally, consider an arbitrary measurable $C_i$. For all $\epsilon > 0$, there exists 
a subset of the circle, $C_i'$ that is a finite, disjoint union of closed intervals
    such that~\cite[Theorem 1.20]{Folland:RealAnalysis}
\begin{equation}
    E[|1(Z \in C_i) - 1(Z \in C_i')|] \le \epsilon.
\end{equation}
Then we have
\ifdefined \JOURNAL 
\begin{align}
    \left \Vert \EE\Brack{\vctZ \mid \vctZ \in C_i}  - \EE\Brack{\vctZ \mid \vctZ \in C_i'} \right \rVert^2 
    &= \left \lVert \EE\Brack{\frac{\vct{Z}1\Paren{\vctZ \in C_i}}{\darc(C_i)} - \frac{\vct{Z}1\Paren{\vctZ \in C_i'}}{\darc(C_i')}} \right \rVert^2 \\
    &\le \EE \Brack{\Paren{\frac{1(Z \in C_i)}{\darc(C_i)} - \frac{1(Z \in C_i')}{\darc(C_i')}}^2} \\
    & \le \frac{\epsilon}{\min^2(\darc(C_i),\darc(C_i'))} +
        \frac{(\darc(C_i) - \darc(C_i'))^2}{\darc(C_i)^2 \darc(C_i)^2} \\
        & \le \frac{\epsilon(1 + \epsilon)}{(\darc(C_i) - \epsilon)^2}. 
\end{align}
\else 
\begin{align}
	&\left \Vert \EE\Brack{\vctZ \mid \vctZ \in C_i}  - \EE\Brack{\vctZ \mid \vctZ \in C_i'} \right \rVert^2 \\
	&= \left \lVert \EE\Brack{\frac{\vct{Z}1\Paren{\vctZ \in C_i}}{\darc(C_i)} - \frac{\vct{Z}1\Paren{\vctZ \in C_i'}}{\darc(C_i')}} \right \rVert^2 \\ 
	&  \le \EE \Brack{\Paren{\frac{1(Z \in C_i)}{\darc(C_i)} - \frac{1(Z \in C_i')}{\darc(C_i')}}^2} \\
	& \le \frac{\epsilon}{\min^2(\darc(C_i),\darc(C_i'))} +
	\frac{(\darc(C_i) - \darc(C_i'))^2}{\darc(C_i)^2 \darc(C_i)^2} \\
	& \le \frac{\epsilon(1 + \epsilon)}{(\darc(C_i) - \epsilon)^2}. 
\end{align}
\fi
Thus, since $C_i'$ satisfies the triangle inequality,
\ifdefined \JOURNAL 
\begin{align}
    \left \Vert \EE\Brack{\vctZ \mid \vctZ \in C_i} \right \rVert & \le
    \left \Vert \EE\Brack{\vctZ \mid \vctZ \in C_i'} \right \rVert + 
    \left \Vert \EE\Brack{\vctZ \mid \vctZ \in C_i} - \EE\Brack{\vctZ \mid \vctZ \in C_i'} \right \rVert \\
     & \le \left|\sinc\Paren{\frac{\darc(C_i')}{2}}\right| + \frac{\sqrt{\epsilon(1-\epsilon)}}{(\darc(C_i) - \epsilon)^2} \\
     & \le \left|\sinc\Paren{\frac{\darc(C_i)-\epsilon}{2}}\right| + \frac{\sqrt{\epsilon(1-\epsilon)}}{(\darc(C_i) - \epsilon)^2}.
\end{align}
\else 
\begin{align}
	&\left \Vert \EE\Brack{\vctZ \mid \vctZ \in C_i} \right \rVert \\
	 & \le
	\left \Vert \EE\Brack{\vctZ \mid \vctZ \in C_i'} \right \rVert + 
	\left \Vert \EE\Brack{\vctZ \mid \vctZ \in C_i} \right.\\
	& \hspace{5mm} \left. - \EE\Brack{\vctZ \mid \vctZ \in C_i'} \right \rVert \\
	& \le \left|\sinc\Paren{\frac{\darc(C_i')}{2}}\right| + \frac{\sqrt{\epsilon(1-\epsilon)}}{(\darc(C_i) - \epsilon)^2} \\
	& \le \left|\sinc\Paren{\frac{\darc(C_i)-\epsilon}{2}}\right| + \frac{\sqrt{\epsilon(1-\epsilon)}}{(\darc(C_i) - \epsilon)^2}.
\end{align}
\fi
Since $\epsilon > 0$ was arbitrary, (\ref{eq:circ:target}) and the theorem follows.

%For fixed $\darc(C_l), \darc(C_r), \sqnrm{\EE\Brack{\vctZ \mid \vctZ \in C_l}}, \sqnrm{\EE\Brack{\vctZ \mid \vctZ \in C_r}}$, $\EE\Brack{\vctZ \mid \vctZ \in C_l}^\top \EE\Brack{\vctZ \mid \vctZ \in C_r}$ is maximized when $C_l$ and $C_r$ are adjacent making $C$ contiguous. 

%    for a given $\darc(C_i)$, $\left \Vert \EE\Brack{\vctZ \mid \vctZ \in C_i} \right \rVert^2$ is maximized if $C_i$ is contiguous. Assume to the contrary that there exists some $C$ such that $C = C_l \cup C_r$ where $C_l$ and $C_r$ are contiguous intervals such that $C_l \cap C_r = \phi$. Apart from $C_l$ and $C_r$, this divides $\cC$ into two more disjoint intervals, $C_1$ and $C_2$. We argue that a rearrangement of the sets results in an encoder that has the same entropy but higher $\left \Vert \EE\Brack{\vctZ \mid \vctZ \in C_i} \right \rVert^2$ and in turn, a lower distortion. 
\end{proof}
\else \begin{proof}[Proof Sketch]  For an encoder $\circenc$ and for $i \in \NN$, we consider quantization cells on the unit circle, $C_i$. The entropy can be written in terms of the Lebesgue measure of the $C_i$, $\mu(C_i)$. The main part of the proof involves bounding the distortion as
	\[ \dist\left( \circenc \right) \geq \sum\limits_i \frac{\mu(C_i)}{2\pi} \Paren{1 - \sinc^2\Paren{\frac{\mu(C_i)}{2}} }  \]
	for an arbitrary measurable $C_i$ with equality holding if each of the $C_i$'s are contiguous arcs.
\end{proof} \fi

%Theorem~\ref{thm:circleED} implies that the optimal quantization cells for the circle source are contiguous arcs. 
%Although Theorem~\ref{thm:circleED} provides a complete characterization of the entropy-distortion tradeoff, it is not computable in the given form. We will now derive computable bounds to the entropy-distortion function. 

\begin{theorem}\label{thm:circleEDLB}
	\ifdefined \JOURNAL 
	\begin{equation}
		\ecirc(D) \geq \sup\limits_{\lambda \geq 0} \inf\limits_{0 < \theta < 2\pi} -\log\Paren{\frac{\theta}{2\pi}} + \lambda \Paren{1-\sinc^2\Paren{\frac{\theta}{2}}} - \lambda D.
	\end{equation}
\else 
\begin{align}
	\ecirc(D) &\geq \sup\limits_{\lambda \geq 0} \inf\limits_{0 < \theta < 2\pi} -\log\Paren{\frac{\theta}{2\pi}} \\
	&+\hspace{5mm} \lambda \Paren{1-\sinc^2\Paren{\frac{\theta}{2}}} - \lambda D.
	\end{align}
\fi 
\label{thm:circ:lower}
\end{theorem}
\ifnum0\ifdefined\EXTENDED 1\fi \ifdefined\JOURNAL 1\fi >0
\begin{proof}
	By weak duality, 
	\ifdefined \JOURNAL 
	\begin{align*}
		\ecirc(D) &\geq \sup\limits_{\lambda \geq 0} \inf_{\substack{\left\{ \theta_i \right\}_{i=1}^{\infty}: \\ \sum_i \theta_i = 2\pi, \\ \theta_i \geq 0 \text{ for all } i}} \sum\limits_{i}^{\infty} -\frac{\theta_i}{2\pi} \log\Paren{\frac{\theta_i}{2\pi}} + \lambda \Paren{\sum\limits_{i=1}^{\infty} \frac{\theta_i}{2\pi} \left( 1- \sinc^2\Paren{\frac{\theta_i}{2}} \right) - D} \\
		&= \sup\limits_{\lambda \geq 0} \inf_{\substack{\left\{ \theta_i \right\}_{i=1}^{\infty}: \\ \sum_i \theta_i = 2\pi, \\ \theta_i \geq 0 \text{ for all } i}} \sum\limits_{i}^{\infty} \frac{\theta_i}{2\pi} \Brack{-\log\Paren{\frac{\theta_i}{2\pi}} + \lambda  \left( 1- \sinc^2\Paren{\frac{\theta_i}{2}} \right) - \lambda D} \\
		&\geq \sup\limits_{\lambda \geq 0} \inf\limits_{0 < \theta < 2\pi} -\log\Paren{\frac{\theta}{2\pi}} + \lambda  \left( 1- \sinc^2\Paren{\frac{\theta}{2}} \right) - \lambda D.
	\end{align*}
	\else
	\begin{align*}
		\ecirc(D) &\geq \sup\limits_{\lambda \geq 0} \inf_{\substack{\left\{ \theta_i \right\}_{i=1}^{\infty}: \\ \sum_i \theta_i = 2\pi, \\ \theta_i \geq 0 \text{ for all } i}} \sum\limits_{i}^{\infty} -\frac{\theta_i}{2\pi} \log\Paren{\frac{\theta_i}{2\pi}} \\
		& \hspace{5mm} + \lambda \Paren{\sum\limits_{i=1}^{\infty} \frac{\theta_i}{2\pi} \left( 1- \sinc^2\Paren{\frac{\theta_i}{2}} \right) - D} \\
		&= \sup\limits_{\lambda \geq 0} \inf_{\substack{\left\{ \theta_i \right\}_{i=1}^{\infty}: \\ \sum_i \theta_i = 2\pi, \\ \theta_i \geq 0 \text{ for all } i}} \sum\limits_{i}^{\infty} \frac{\theta_i}{2\pi} \left[ -\log\Paren{\frac{\theta_i}{2\pi}} \right. \\
		&\hspace{5mm}\left. + \lambda  \left( 1- \sinc^2\Paren{\frac{\theta_i}{2}} \right) - \lambda D\right] \\
		&\geq \sup\limits_{\lambda \geq 0} \inf\limits_{0 < \theta < 2\pi} -\log\Paren{\frac{\theta}{2\pi}} \\ 
		&\hspace{5mm}+ \lambda  \left( 1- \sinc^2\Paren{\frac{\theta}{2}} \right) - \lambda D.
	\end{align*} 
	\fi 
	\end{proof}
\else \fi

%The above lower bound implies a lower bound of $\frac{1}{2}$ on the circle's $\hslope$. 

%ANN-based compressors minimize the Lagrangian of the entropy-distortion function using stochastic optimization methods. A lower bound similar in spirit to Theorem~\ref{thm:circleEDLB} also holds for minimizing the Lagrangian. 
%
%\ifdefined \JOURNAL
%\begin{Corollary}
%For $\lambda \geq 0$, 
%\[ \inf_{\fenc} H\Paren{\fenc\Paren{\vctZ}} + \lambda \EE\Brack{\left \lVert \vctZ - \EE\Brack{\vctZ \mid \fenc\Paren{\vctZ}} \right \rVert^2} \geq  \inf\limits_{0 < \theta < 2\pi} -\log\Paren{\frac{\theta}{2\pi}} + \lambda \Paren{1-\sinc^2\Paren{\frac{\theta}{2}}}.\]
%\end{Corollary}
%\else 
%\begin{Corollary}
%	For $\lambda \geq 0$, 
%	\begin{align*}
%		&\inf_{\fenc} H\Paren{\fenc\Paren{\vctZ}} + \lambda \EE\Brack{\left \lVert \vctZ - \EE\Brack{\vctZ \mid \fenc\Paren{\vctZ}} \right \rVert^2} \\
%		&\geq  \inf\limits_{0 < \theta < 2\pi} -\log\Paren{\frac{\theta}{2\pi}} + \lambda \Paren{1-\sinc^2\Paren{\frac{\theta}{2}}}.\end{align*}
%\end{Corollary}
%
%\fi

The lower bound in Theorem~\ref{thm:circ:lower} is illustrated in Fig.~\ref{fig:c_RD} (``lower bound'').  An upper bound can be obtained by partitioning the circle into arcs with a biuniform size distribution (``achievable for biuniform'' intervals in Fig.~\ref{fig:c_RD}). Note that these bounds essentially coincide at high-SNR.
%Note that this is the same upper bound as obtained by~\cite{KolianderPRH16}. 

%Like the ramp, for rates of the form $\log K$ where $K\geq 2$ is an integer, an achievable scheme partitions the circle into $K$ equal arcs. For rates between $\log K$ and $\log K+1$, an achievable scheme is obtained by dividing the circle into biuniform arcs, where one arc has length $2 \pi \eps$ and the other $K$ arcs have length $\frac{2 \pi (1-\eps)}{K}$. From Figure~\ref{fig:c_RD} we see that the upper and lower bounds to the optimal tradeoff are close.

%The encoder assigns a point on the circle to the corresponding index of the arc. The decoder outputs the conditional mean of the distribution restricted to the arc. Since the arcs are equal, the rate is $\log K$. The distortion is 
%\ifdefined \JOURNAL 
%\begin{align*}
%	\sum\limits_{i=1}^{K} \frac{1}{K}\EE\Brack{ \left \lVert \vctZ - \EE\Brack{\vctZ \mid \vctZ \in C_i} \right \rVert^2 \mid \vctZ \in C_i} = 1 - \frac{1}{K}\sum\limits_{i=1}^{K}\left \lVert \EE\Brack{\vctZ \mid \vctZ \in C_i} \right \rVert^2 = 1- \sinc^2\Paren{\frac{\pi}{K}}.
%\end{align*}
%\else
%\begin{align*}
%	&\sum\limits_{i=1}^{K} \frac{1}{K}\EE\Brack{ \left \lVert \vctZ - \EE\Brack{\vctZ \mid \vctZ \in C_i} \right \rVert^2 \mid \vctZ \in C_i} \\
%	& = 1 - \frac{1}{K}\sum\limits_{i=1}^{K}\left \lVert \EE\Brack{\vctZ \mid \vctZ \in C_i} \right \rVert^2 = 1- \sinc^2\Paren{\frac{\pi}{K}}.
%\end{align*}
% \fi 

\ifdefined\JOURNAL
\begin{figure}
	\centering
	\includegraphics[width=\linewidth]{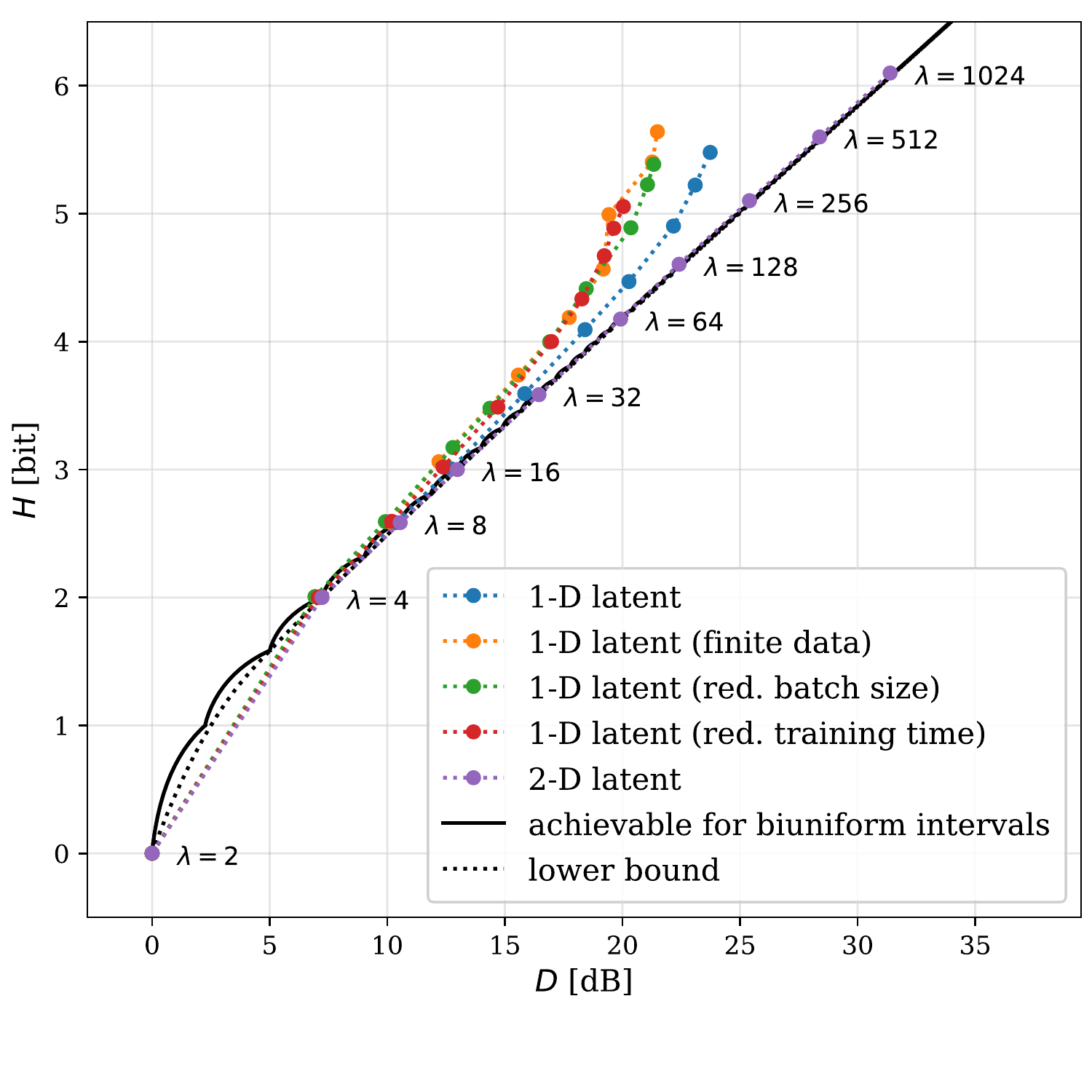}
	\caption{Circle: Entropy-Distortion tradeoff of ANN-based compressors with latent dimensions $1$ and $2$ along with lower bound and upper bound on optimal tradeoff.}\label{fig:c_RD}
\end{figure}
\else 
\begin{figure}[!tb]
	\centering
	\includegraphics[trim = 0 30 0 0 , clip, width=\linewidth]{sphere_rd}
	\caption{Circle: Entropy-Distortion tradeoff of ANN-based compressors with latent dimensions $1$ and $2$ along with lower bound and upper bound on optimal tradeoff.}\label{fig:c_RD}
\end{figure}
\fi

\subsection{ANN Performance}

\ifdefined \JOURNAL \else To compress a source of dimension $\sdim$ using a latent dimension of $\cdim$, a stochastically-trained ANN-based compressor consists of an analysis transform $\anl:\RR^{\sdim} \mapsto \RR^{\cdim}$, a synthesis transform $\syn:\RR^{\cdim} \mapsto \RR^{\sdim}$, a quantizer and an entropy model that is factorized across each dimension of the codeword. The analysis and synthesis transforms are fully-connected feedforward neural networks with $2$ hidden layers containing $100$ neurons each. The quantization operation is not differentiable,
and therefore during training we replace it with a differentiable proxy that varies over the course of the
training process, beginning with dithered quantization and ending with the hard quantizer that is used
at test time \cite{AgustssonT20}. During training, a $\sdim$-dimensional vector is fed into the analysis transform to obtain a $\cdim$-dimensional latent vector. The quantization-proxy is then applied to the latent vector, which is then fed to the synthesis transform to obtain the $\sdim$-dimensional reconstruction. The entropy model is a feedforward neural network that computes the entropy, $E$ of the quantized latents. Distortion $D$ is the mean-squared error between the reconstruction and the input vector. The Lagrangian $E+\lambda D$ is stochastically minimized over the trainable parameters of the ANN-based compressor using Adam \cite{KingmaBBL15}. We sweep across different values of $\lambda$ to obtain points on the lower convex hull of the ANN-compressor's entropy-distortion tradeoff. We remark that the neural-network architecture and entropy model are similar to the ANN-based compressor trained for the sawbridge process in \cite{WaBa:Sawbridge:Long}. 
\fi
Since the circle is described by the scalar random variable
$\theta$, we take the latent dimension $\cdim = 1$. The 
resulting performance is shown in Fig.~\ref{fig:c_RD}. We see that 
the performance is suboptimal at high-SNR.

%The performance is suboptimal for both 1-D and 2-D latent spaces.
%In both cases, the issue is that the analysis transform seeks
%to implement a function with a step discontinuity (or 
%at least a function with a large Lipschitz constant), but
%has difficulty learning this function during the stochastic training
%process.
%The details, however, are slightly different between the
%1-D and 2-D cases.

\begin{figure*}[!htb]
%	\begin{subfigure}{.32\textwidth}
%		\centering
%		\includegraphics[trim =  0 10 0 0, clip, scale=0.34]{c_1D_latent_low_rate}
%		\caption{}\label{fig:c_anl_low}
%	\end{subfigure}%
    \begin{center}
	\begin{subfigure}{0.48\textwidth}
		\centering
		\includegraphics[trim = 10 10 0 0, clip, scale=0.39]{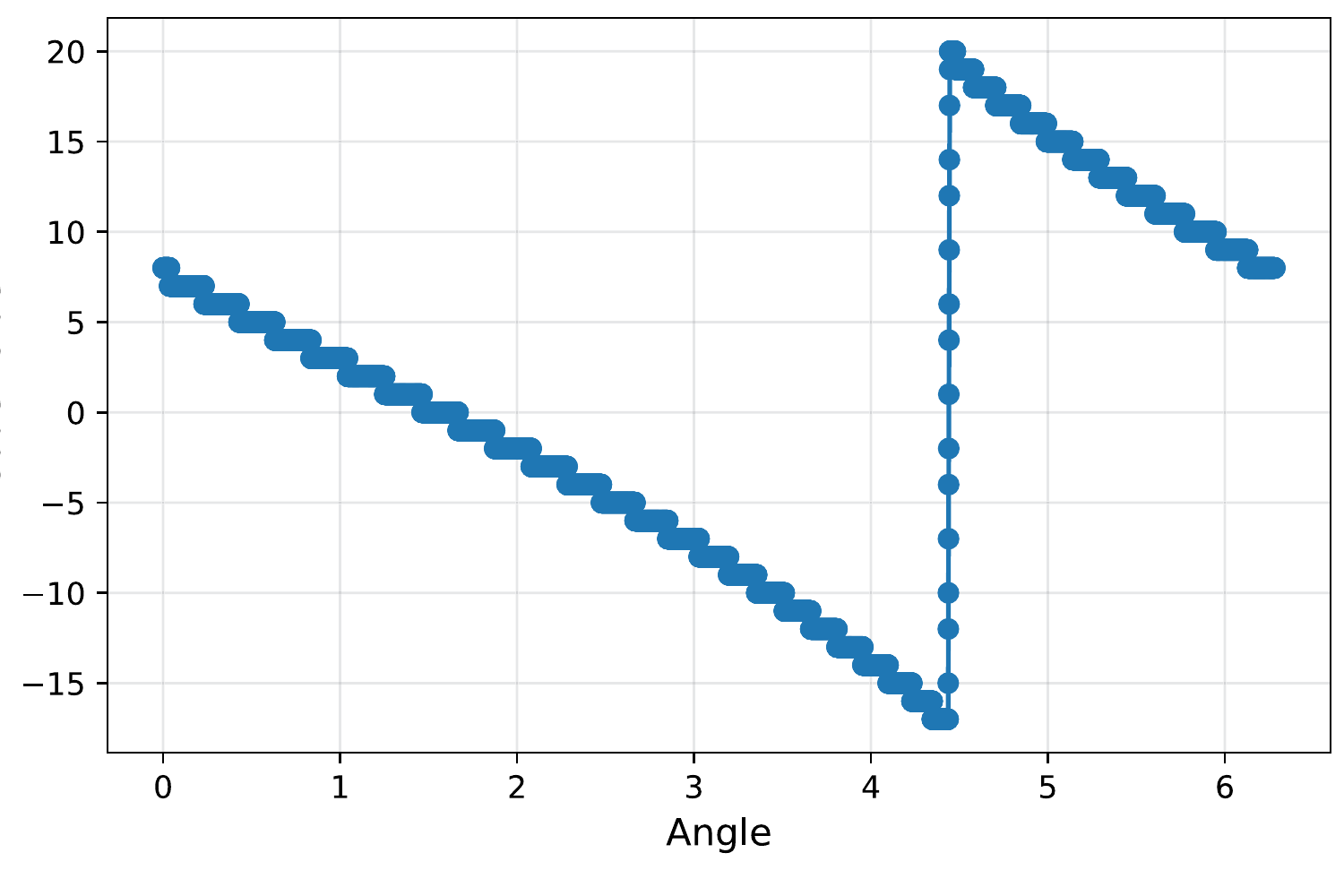}
		\caption{}\label{fig:c_anl}
	\end{subfigure}
	\begin{subfigure}{0.48\textwidth}
		\centering
		\includegraphics[trim = 10 10 0 0, clip, scale=0.39]{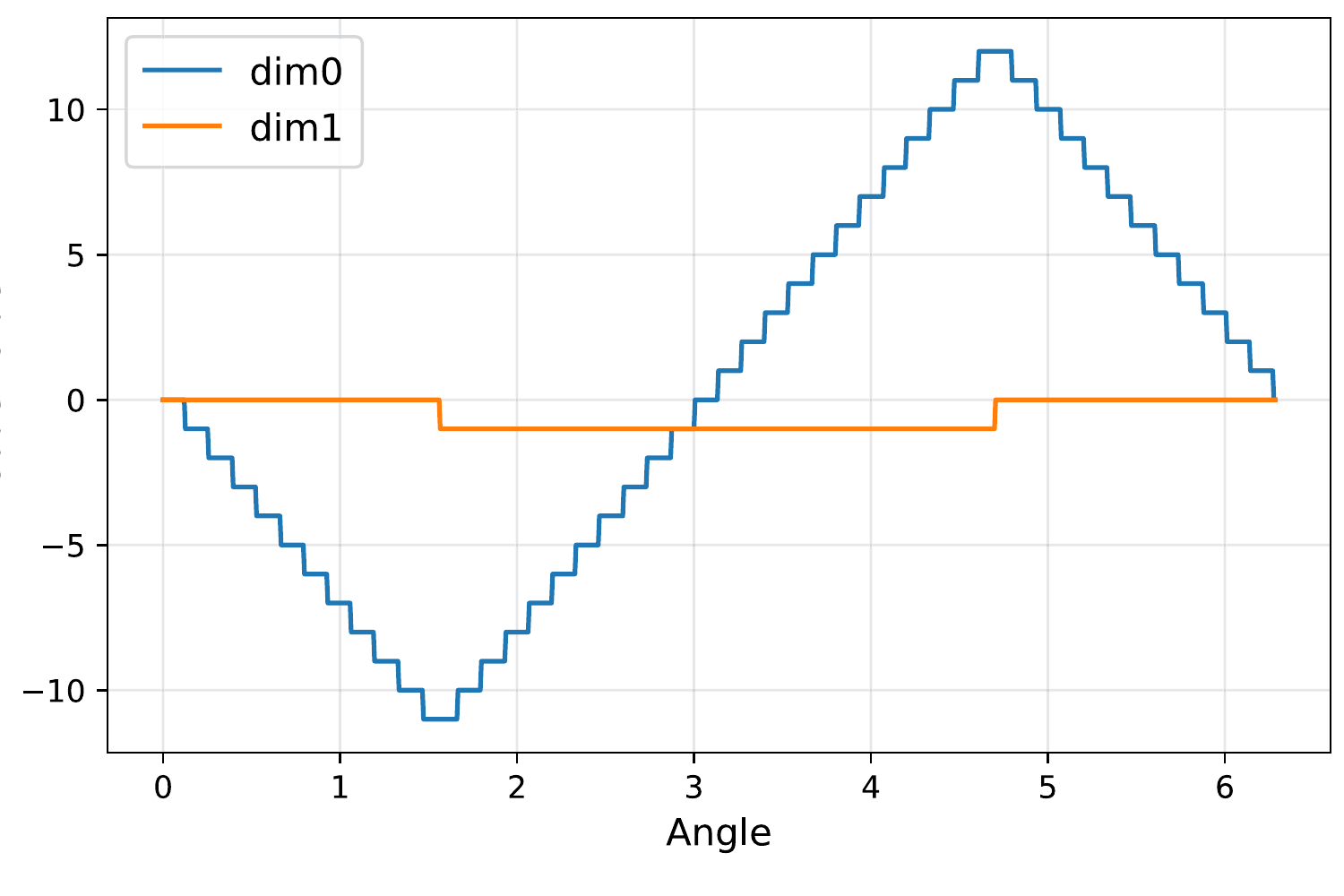}
		\caption{}\label{fig:c_anl_2D}
	\end{subfigure}
    \end{center}
	\caption{
       % (a) Codeword value vs angle for $\lambda=4$ (on optimal tradeoff) and $1$D latent. The unquantized plot is obtained by passing the output of $\anl$ through $\syn$ without quantization. The analysis transform is sufficiently steep because of quantization. 
        (a) Quantized encoder output vs. angle $\theta$ for $\lambda=512$ and $1$-D latent (away from optimal tradeoff). The analysis transform is not sufficiently steep. (c) Quantized encoder output vs. angle $\theta$ for $\lambda=512$ and $2$-D latents (on optimal tradeoff). 2-D latents are discussed in Section~\ref{sec:2D}.}
\end{figure*}

\ifdefined\JOURNAL 
\subsubsection{Lipschitz-Limiting Condition}
We now highlight the fundamental difficulty for the suboptimal performance of ANN-based compressors on the circle source. Intuition suggests, and the proofs of Theorems~\ref{thm:circleED} 
 and~\ref{thm:circleEDLB}
 essentially confirm, that an optimal scheme for compressing the
 circle is to quantize the angle $\theta$ into contiguous cells,
 all but (at most) one of which are the same size. 
 Thus we would like the analysis transform to extract the angle
 $\theta$ from the realization of the circle, i.e., to 
 implement $\mathrm{atan2(\vct{Z})}$ or some scaled and 
 shifted version
 thereof. 
 
 While the function $\theta \mapsto \vct{Z} = \left(\cos(2\pi\theta),\sin(2\pi\theta)\right)$ is continuous, the issue is that the inverse function $\vct{Z} \mapsto \theta$
 is not continuous over the circle since the unit circle and the line segment $\left[0,1\right)$ are not homeomorphic \cite[Section 18, Ex. 6]{Munkres00}. However the 
 analysis transform, being a composition of alternating linear and nonlinear continuous and differentiable functions, must be continuous (and indeed differentiable). 
 We therefore say that the source exhibits a ``Lipschitz-limiting condition". Note that this condition is irrespective of the choice of network architecture and choice of non-linearities in the analysis and synthesis transform. 
 
\else
To see why, note that 
intuition suggests, and the proofs of Theorems~\ref{thm:circleED} 
and~\ref{thm:circleEDLB}
essentially confirm, that an optimal scheme for compressing the
circle is to quantize the angle $\theta$ into contiguous cells,
all but (at most) one of which are the same size. 
Thus we would like the analysis transform to extract the angle
$\theta$ from the realization of the circle, i.e., to 
implement $\mathrm{atan2(\vct{Z})}$ or some scaled and 
shifted version
thereof. The issue is that the function $\vct{Z} \mapsto \theta$
is not continuous over the circle, while the 
analysis transform must be continuous (and indeed differentiable)
by construction.
\fi
%More precisely, the analysis transform of an ANN-based compressor, whose code dimension is $1$, is a continuous and bounded mapping $\anl:\RR^2\mapsto \RR$,
%which we assume is not constant. 
%Consider $\anl$ as a function of $\theta$, $h : \left[0,2\pi \right] \mapsto \left[0,1\right]$ where $h(\theta) = \anl(\left[ \cos \theta, \sin \theta \right])$ and both $0$ and $1$ are attained for some $\theta \in \left[0,2\pi\right)$. 
%By continuity of $\anl$, $h(0) = h(2\pi)$. By the extreme value theorem, there exists $\theta_{max}$ and $\theta_{min}$ such that $h(\theta_{max}) = \sup_\theta h(\theta) $ and $h(\theta_{min}) = \inf_\theta h(\theta)$. 
%Without loss of generality, we may assume that 
%$\sup_\theta h(\theta) = 1$ and $\inf_\theta h(\theta) = 0.$
%Since $h(0) = h(2\pi)$, for every $z \in \left[0,1\right]$, there exist at least two angles $\theta_{major}, \theta_{minor}$ such that $h(\theta_{major}) = h(\theta_{minor}) = z$, i.e., for every point in the code space, the analysis transform maps at least two distinct angles to the same point. 

\ifdefined\JOURNAL The Lipschitz-limiting condition \else This \fi is confirmed in Fig.~\ref{fig:c_anl}, which 
shows the quantized output of the trained analysis transform $\anl$ 
as a function of the angle $\theta$ at high SNR. We see that the
analysis transform attempts to implement a discontinuity around
4.5 radians, but the function is insufficiently steep, so it
passes through various intermediate quantization levels on
its way from its minimum value to its maximum value. This
creates an identifiability problem at the decoder, in that
certain quantizer outputs can be caused by two values of
$\theta$, one in the decreasing part of the function and
one in the increasing part. Of course, the location of the
discontinuity (4.5 radians in this case) is arbitrary and
will be different if the network is retrained.  \ifdefined \JOURNAL Since bulk of the distortion is due to the source realizations in the ``steep" portion of the analysis transform

At low SNR, the distortion accruing from \ifdefined \JOURNAL the Lipschitz-limiting condition \else this lack of invertibility \fi
is negligible compared to the distortion arising from
the quantization process. At high rates,
it dominates, and the performance is off the optimal
entropy-distortion curve. 
The extent of the suboptimality is 
determined by how steep the analysis transform can be made, which in 
turn is controlled by the training process.
In order for the training process to make the ``steep''
portion of the analysis transform steeper, we require
source realizations from the range over which the function
is steep to be present in the batch. Smaller batch sizes are less 
likely to include such points, and thus reducing the batch size 
leads to a larger gap from optimality (Fig.~\ref{fig:c_RD}).
Fixing a training data set (instead of drawing fresh samples
at each iteration) has a similar effect, because once the
steep portion of the function falls entirely between two 
training points, the training process has no incentive
to make it steeper (Fig.~\ref{fig:c_RD}). In Appendix~\ref{app:varblowup} we provide an explanation as to why stochastic minimization algorithms that sample points uniformly from the source do not achieve the optimal transforms.

\else
At low SNR, the distortion accruing from this lack of invertibility
is negligible compared to the distortion arising from
the quantization process. At high rates,
it dominates, and the performance is off the optimal
entropy-distortion curve. 
The extent of the suboptimality is 
determined by how steep the analysis transform can be made, which in 
turn is controlled by the training process.
In order for the training process to make the ``steep''
portion of the analysis transform steeper, we require
source realizations from the range over which the function
is steep to be present in the batch. Smaller batch sizes are less 
likely to include such points, and thus reducing the batch size 
leads to a larger gap from optimality (Fig.~\ref{fig:c_RD}).
Fixing a training data set (instead of drawing fresh samples
at each iteration) has a similar effect, because once the
steep portion of the function falls entirely between two 
training points, the training process has no incentive
to make it steeper (Fig.~\ref{fig:c_RD}).
\fi
This issue is not unique to the circle in the Euclidean
plane. Indeed, it can arise in more complex sources 
whose support has a circular structure, such as the
\emph{ramp} process to which we turn next.

\section{The Ramp}\label{sec:ramp}

Consider the following process.
\begin{equation}\label{eq:rampdef}
    \rmp_t \eqdef \left[ \left( t + \phase \right) \text{ mod } 1  \right] - \frac{1}{2},
\end{equation}
where $\phase \sim \text{Unif}\left[ 0,1\right]$. 
We call this the \emph{ramp} process and 
$\phase$ the \emph{phase}. We are interested
in this process as a model of low-dimensional
structure in high-dimensional spaces: on the
one hand, the set of source realizations has
infinite linear span; on the other hand, the
realization is completely determined by the
scalar random variable $\phase$.
Note that $\phase=1$ and $\phase=0$ yield 
identical realizations of the ramp. Thus the
set of realizations forms a circle in function
space in some sense.
%We say that the ramp has \emph{circular symmetry} and that the 
%phase is a \emph{circular latent variable}. 
This process is similar in some important
respects to the ``sawbridge'' process considered in
an earlier work~\cite{WaBa:Sawbridge:Long}. But whereas the 
sawbridge is optimally compressed by the ANN-based
architecture under study, we shall see that the
ramp is not.

\ifdefined\EXTENDED
Since realizations of the ramp process are in one-to-one
correspondence with realizations of $\phase$, it
is natural to compress the ramp process by 
quantizing the $\phase$. The next theorem confirms
that this is indeed optimal.
\fi

%TODO: Add figure for ramp realizationss

%The ramp can be thought of as a simplification of the stationary sawbridge. To see, this first note that an alternative representation of the nonstationary sawbridge is 
%\[ \saw_t = \left( t - \drop_{DC} - \frac{1}{2} \right) \text{ mod }1 + \drop_{DC} - \frac{1}{2},\]
%where $\drop_{DC} \sim \text{Unif}\left[0,1\right]$. This implies that the stationary sawbridge can be represented as
%\[ \stsaw_t = \left( \Paren{t + \phase} \text{mod }1 - \drop_{DC} - \frac{1}{2} \right) \text{ mod }1 + \drop_{DC} - \frac{1}{2}.\]
%The ramp can therefore be thought of as subtracting the DC component from the stationary sawbridge and isolating the phase information from the stationary sawbridge. The variance of the ramp is $\frac{1}{12}$.

%\begin{proposition}
%	If $h$ is continuous and bounded, then by  the extreme value theorem, there exists $\theta_{max}$ and $\theta_{min}$ such that $h(\theta_{max}) = 1$ and $h(\theta_{min}) = 0$. Since $h(0) = h(2\pi)$, for every $z \in \left[0,1\right]$, 
%\end{proposition}

\subsection{The Optimal Tradeoff}\label{subsec:rampRD}

\begin{theorem}\label{thm:rampED}
	For the ramp, if $D \geq \frac{1}{12}$, $\eramp(D) = 0$. If $0 < D < \frac{1}{12}$, then
    \begin{align}
        \eramp(D) = \inf\limits_{\left\{ p_i \right\}_{i=1}^{\infty}} \quad -&\sum\limits_{i = 1}^{\infty} p_i \log p_i  \label{eq:rampopt1} \\
        \textrm{s.t. } \quad& \sum\limits_{i=1}^{\infty} \frac{p_i^2 (2 - p_i)}{12} \leq D, \label{eq:rampopt2} \\
	& \sum\limits_{i=1}^{\infty} p_i = 1, p_i \geq 0 \text{ for all i}. \label{eq:rampopt3}
    \end{align}
\end{theorem}

\ifnum0\ifdefined\EXTENDED 1\fi \ifdefined\JOURNAL 1\fi >0
\begin{proof}
	Let $\rampenc: \lebtwo \mapsto \NN$ be an encoder. 
    Each value of the phase variable defines a unique realization. Therefore, define $\rset_i = \rampenc^{-1} \left( i\right) $ where
    \begin{equation}
        \label{eq:ramp:Sdef}
       \rampenc^{-1} \left( i\right) \eqdef \left \lbrace v \in \left[0,1\right]: \rampenc\left([ (t+v) \text{ mod} 1] - 0.5\right)= i \right \rbrace.
\end{equation}
    Let $\leb$ be the Lebesgue measure.
    The optimal reconstruction is $t \mapsto \EE\left[ \rmp_t \mid \rampenc(\rmp_t)\right]$. 
    Thus the entropy and distortion of the encoder-decoder pair is given by 
	\begin{align*}
	\ent\left(\rampenc\right) &=   -\sum_i \leb(\rset_i) \log \left( \leb(\rset_i)\right), \\
	\dist\left( \rampenc \right) & =  \EE\left[ \int_0^1 \left( \rmp_t - \EE\left[ \rmp_t \mid \rampenc(\rmp_t)\right] \right)^2 dt \right].  
	\end{align*}

The distortion can be expressed as 
\ifdefined \JOURNAL 
\begin{align}
        \dist\left( \rampenc \right)=&\EE\left[ \int_0^1 \left( \rmp_t - \EE\left[ \rmp_t \mid \rampenc(\rmp_t)\right] \right)^2 dt \right] \\
		&= \sum_i \leb\left( \rset_i \right) \EE\left[ \int_0^1 \left( \rmp_t - \EE\left[ \rmp_t \mid \phase \in \rset_i \right] \right)^2 dt \middle| \phase \in \rset_i \right] \\
		&= \sum_i \leb\left( \rset_i \right) \left( \EE\left[ \int_0^1 \rmp_t^2 dt \mid \phase \in \rset_i \right] - \int_0^1 \left( \EE\left[ \rmp_t \middle| \phase \in \rset_i \right] \right)^2 \; dt \right) \\
 &  = \sum_i \leb\left( \rset_i \right) \left( \frac{1}{12} - \int_0^1 \left( \EE\left[ \rmp_t \mid \phase \in \rset_i \right] \right)^2 dt \right).
    \label{eq:ramp:expansion}
\end{align}
\else 
\begin{align}
	&\EE\left[ \int_0^1 \left( \rmp_t - \EE\left[ \rmp_t \mid \rampenc(\rmp_t)\right] \right)^2 dt \right] \\
	&= \sum_i \leb\left( \rset_i \right) \EE\left[ \int_0^1 \left( \rmp_t - \EE\left[ \rmp_t \mid \phase \in \rset_i \right] \right)^2 dt \middle| \phase \in \rset_i \right] \\
	&= \sum_i \leb\left( \rset_i \right) \left( \EE\left[ \int_0^1 \rmp_t^2 dt \mid \phase \in \rset_i \right] - \right. \\
	&\left. \hspace{40mm} \int_0^1 \left( \EE\left[ \rmp_t \middle| \phase \in \rset_i \right] \right)^2 \; dt \right) \\
	&  = \sum_i \leb\left( \rset_i \right) \left( \frac{1}{12} - \int_0^1 \left( \EE\left[ \rmp_t \mid \phase \in \rset_i \right] \right)^2 dt \right).
	\label{eq:ramp:expansion}
\end{align}
\fi
where the last line follows because,
irrespective of the phase, we have $\int_0^1 \rmp_t^2 \; dt =  \frac{1}{12}$.

We will show that, for any measurable $\rset \subset [0,1]$, we have
\begin{equation}\label{eq:rampineq}
\int_0^1 \left( \EE\left[ \rmp_t \mid \phase \in \rset \right] \right)^2 dt \leq \frac{1}{3} \Paren{\frac{1-\mu\Paren{\rset}}{2}}^2.
\end{equation}
%Note that both sides of the inequality are invertible with
%respect to cyclic shifts of $\rset$ of the form $\rset \rightarrow
%s + \rset \mod 1$. 
Write $y_t = E[\rmp_t| \phase \in \rset]$. Then we have
% TODO: need consistent notation for indicator

\ifdefined \JOURNAL 
\begin{align}
    \label{eq:ramp:seeshift}
     &  \EE\left[(t + V \mod 1) - 
        \frac{1}{2} \middle| V \in \rset\right] \\
        \nonumber
    &= \EE \left[ \Paren{t + \phase - \frac{3}{2}} 1(t + \phase > 1) 
         + \Paren{t + \phase - \frac{1}{2}} 1(t + \phase \le 1)
             \middle| \phase \in \rset\right] \\ 
             \nonumber
             & = t + \EE[\phase|\phase \in \rset] -
             \PP\Paren{t + \phase > 1\mid\phase \in S} - \frac{1}{2} \\
             \label{eq:rampident}
             & = t + \EE[\phase|\phase \in \rset] -
             \frac{\mu\Paren{\rset \cap [1-t,1]}}{\mu(\rset)}
              - \frac{1}{2}.
\end{align}
\else 
\begin{align}
	\label{eq:ramp:seeshift}
	&  \EE\left[(t + V \mod 1) - 
	\frac{1}{2} \middle| V \in \rset\right] \\
	\nonumber
	& = \EE \left[ \Paren{t + \phase - \frac{3}{2}} 1(t + \phase > 1) 
	\right. \\ \nonumber 
	&\hspace{5mm} + \left. \Paren{t + \phase - \frac{1}{2}} 1(t + \phase \le 1)
	\middle| \phase \in \rset\right] \\ 
	\nonumber
	& = t + \EE[\phase|\phase \in \rset] -
	\PP\Paren{t + \phase > 1\mid\phase \in S} - \frac{1}{2} \\
	\label{eq:rampident}
	& = t + \EE[\phase|\phase \in \rset] -
	\frac{\mu\Paren{\rset \cap [1-t,1]}}{\mu(\rset)}
	- \frac{1}{2}.
\end{align}
\fi

From \eqref{eq:ramp:seeshift} we 
see that the left-hand side of \eqref{eq:rampineq} is
invariant with respect to cyclic rotation of $\rset$ of the 
form $\rset \mapsto s + \rset \mod 1$. Evidently the right-hand
side of \eqref{eq:rampineq} is also invariant with respect to such a
shift. Thus to show \eqref{eq:rampineq} we may cyclically rotate $\rset$
as convenient.

First, 
suppose that $S$ can be cyclicly rotated to obtain an interval
of the form $[0,s]$. In this case \eqref{eq:rampident} reads
\begin{equation}
    y_t = \begin{cases}
        t\left(1 - \frac{1}{s}\right) + \frac{s}{2} - \frac{3}{2}
            + \frac{1}{s} & \text{if $t > 1-s$} \\
        t -\frac{1 - \mu\Paren{\rset}}{2} &
           \text{if $t \le 1-s$},
       \end{cases}
\end{equation}
and thus we can explicitly compute
\begin{equation}
    \label{eq:rampint}
    \int_0^1 y_t^2 dt = \frac{1}{3} \Paren{\frac{1 - \mu(\rset)}{2}}^2.
\end{equation}
and equality in \eqref{eq:rampineq} holds in this case. More generally,
suppose that $\rset$ is a finite union of closed, disjoint
intervals, $\rset_0,\ldots,\rset_{n-1}$ such that all points in
$\rset_0$ are less than those in $\rset_1$, etc. Due to the
rotational invariance of \eqref{eq:rampineq}, we can assume that
$\min \rset_0 = 0$ and $\max \rset_{n-1} < 1$. 
Define
\begin{align}
    s_i & = \min \rset_i \quad i = 0,\ldots, n - 1 \\
    s_n & = 1 \\
    t_i & = s - s_{n-i} \quad i = 0,\ldots, n.
\end{align}
Note that for each $i$, from \eqref{eq:rampident}, over the interval
$[t_i,t_{i+1}]$, $y_t$ is linearly increasing and then
decreasing.  Thus for $t \in [t_i,t_{i+1}]$, 
\begin{equation}
    y_t \le y_{t_i} + \frac{t - t_i}{t_{i+1} - t_i}(y_{t_{i+1}} - 
        y_{t_i}) \label{eq:ramp:keyinterval}
\end{equation}
and as a result
\begin{equation}
    \int_{t_i}^{t_{i+1}}  y_t \; dt
    \le \frac{y_{t_i} + y_{t_{i+1}}}{2} [t_{i+1} - t_i].
\end{equation}

Since every realization of $J_t$ integrates to zero, we have
\ifdefined\JOURNAL
\begin{equation}
0 = \EE\left[\int_0^1 J_t dt \middle| \phase \in \rset \right]
  = \int_0^1 \EE[J_t|\phase \in \rset] dt = \int_0^1 y_t dt = 
  \sum_{i = 0}^{n-1} \int_{s_i}^{s_{i+1}} y_t dt 
   \le
   \sum_{i = 1}^n \frac{y_{t_i} + y_{t_{i+1}}}{2} \left[ t_{i+1}-t_i \right].
\end{equation}
\else
\begin{align*}
0 &= \EE\left[\int_0^1 J_t dt \middle| \phase \in \rset \right] \\
&= \int_0^1 \EE[J_t|\phase \in \rset] dt \\
&= \int_0^1 y_t dt\\
& = \sum_{i = 0}^{n-1} \int_{s_i}^{s_{i+1}} y_t dt\\ 
&\le \sum_{i = 1}^n \frac{y_{t_i} + y_{t_{i+1}}}{2} \left[ t_{i+1}-t_i \right].
\end{align*}
\fi

It follows that there exists $i$ such that $y_{t_i} + y_{t_{i+1}} > 0$.
By rotational invariance, we can assume that $i = 0$. Consider
the modified set $\tilde{\rset} = \tilde{\rset}_0
\cup \cdots \cup \tilde{\rset}_{n-1}$ wherein 
$\tilde{\rset}_i = \rset_i$ for $i = 0,\ldots,n-2$ and
$\tilde{\rset}_{n-1}$ is a closed interval satisfying
$\mu(\tilde{\rset}_{n-1}) = \mu(\rset_{n-1})$ and
$\max \tilde{\rset}_{n-1} = 1$.
In words, we shift $\rset_{n-1}$ to the right
  so that its right end-point is 1. Define
  \begin{equation}
      \tilde{y}_t = \EE[J_t|\phase \in \tilde{\rset}]
  \end{equation}
  and note that, from \eqref{eq:rampident}, we have
  \begin{equation}
      \tilde{y}_t = \begin{cases}
          y_0 + y_{t_1} - y_{t_1 - t} & \text{if $t \in [0,t_1]$} \\
           y_t & \text{if $t \in (t_1,1]$}.
      \end{cases}
  \end{equation}
  From this it follows that the norm of $\tilde{y}_t$ dominates
  that of $y_t$:
 \ifdefined \JOURNAL
  \begin{align}
      \int_{0}^1 \tilde{y}_t^2 dt 
      & = \int_{0}^{t_1} \Paren{y_0 + y_{t_1} - y_{t_1 - t}}^2 dt  +
         \int_{t_1}^{1} y_t^2 dt \\
         & = \int_0^1 y_t^2 dt + \Paren{y_0 + y_{t_1}}^2 t_1
           - 2(y_0 + y_{t_1}) \int_{0}^{t_1} y_t dt \\
        & \ge \int_0^1 y_t^2 dt,
   \end{align}
\else
  \begin{align}
	\int_{0}^1 \tilde{y}_t^2 dt 
	& = \int_{0}^{t_1} \Paren{y_0 + y_{t_1} - y_{t_1 - t}}^2 dt  +
	\int_{t_1}^{1} y_t^2 dt \\
	& = \int_0^1 y_t^2 dt + \Paren{y_0 + y_{t_1}}^2 t_1 \\
	& \hspace{5mm}- 2(y_0 + y_{t_1}) \int_{0}^{t_1} y_t dt \\
	& \ge \int_0^1 y_t^2 dt,
\end{align}
\fi
  where the final equality follows from \eqref{eq:ramp:keyinterval}
  and the assumption that $y_{t_i} + y_{t_{i+1}} > 0$.
  Now $\tilde{\rset}$ can be written as a union of $n-1$ disjoint
  closed intervals, since $\tilde{\rset}_0$ and $\tilde{\rset}_{n-1}$
  are effectively conjoined.
   Repeating this process until we arrive
   at a single interval and then applying \eqref{eq:rampint} gives
   \begin{equation}
       \int_0^1 \EE[J_t | \phase \in \rset]^2 dt \le
       \frac{1}{3} \Paren{\frac{1 - \mu(\rset)}{2}}^2
   \end{equation}
   for any $\rset$ that is a disjoint union of closed intervals.
   Finally, consider an arbitrary (measurable) $\rset \subset [0,1]$. 
   Then for any
   $\epsilon > 0$, there exists a $\tilde{\rset} \subset [0,1]$
   that is finite union of disjoint closed intervals such that
   \begin{equation}
       \mu\Paren{(\rset \backslash \tilde{\rset}) \cup
       (\tilde{\rset} \backslash \rset)} 
   \end{equation}
   can be made arbitrarily small~\cite[Theorem 1.20]{Folland:RealAnalysis}, and in particular for any
   $\epsilon$ we can have
   \begin{align}
       \label{eq:ramp:close1}
       \left|\frac{\EE[\phase \cdot 1(\phase \in \rset)]}{\PP(\phase \in \rset)}
       - \frac{\EE[\phase \cdot 1(\phase \in \tilde{\rset})]}{\PP(\phase 
       \in \tilde{\rset})}\right| & \le \epsilon \\
       \label{eq:ramp:close2}
       \left|\frac{\mu(\rset \cap [1-t,1])}{\mu(\rset)} -
       \frac{\mu(\tilde{\rset} \cap [1-t,1])}{\mu(\tilde{\rset})}
          \right| & \le \epsilon.
   \end{align}
   Define the two conditional means
   \begin{align*}
       y_t & = \EE[J_t | \phase \in \rset] \\
       \tilde{y}_t & = \EE[J_t | \phase \in \tilde{\rset}].
   \end{align*}
   Then from \eqref{eq:ramp:close1} and \eqref{eq:ramp:close2} we have, for each $t$,
   \begin{equation*}
       y_t \le \tilde{y}_t + 2\epsilon,
   \end{equation*}
% Thus we obtain the bound
% \ifdefined \JOURNAL 
% \begin{align}
%     \int_0^1 y_t^2 dt & \le \left( \frac{\mu(\tilde{\rset})}
% {\mu(\tilde{S}) - 2 \epsilon} \right) \int_0^1 \tilde{y}_t^2 dt + 
%     \Paren{\frac{\epsilon(4+3\mu(\rset))}{\mu(\rset)^2-\eps^2}}^2 \\
%     & \le \frac{1}{3} \cdot \left( \frac{\mu(\tilde{\rset})
%           + \epsilon}{\mu(\tilde{\rset}) - 
%       2\epsilon}\right)^2\frac{(1 - \mu(\tilde{\rset}))^2}{2} +
%          \Paren{\frac{\epsilon(4+3\mu(\rset))}{\mu(\rset)^2-\eps^2}}^2 \\
%          & \le \frac{1}{3} \cdot \left(\frac{\mu(\tilde{\rset})
%              + \epsilon}{\mu(\tilde{\rset}) - 2\epsilon}\right)^2
%               \frac{(1 - \mu(\rset) + \epsilon)^2}{2}
%           +  \Paren{\frac{\epsilon(4+3\mu(\rset))}{\mu(\rset)^2-\eps^2}}^2,
% \end{align}
%\else
%\begin{align}
%	\int_0^1 y_t^2 dt & \le \left( \frac{\mu(\tilde{\rset})}
%	{\mu(\tilde{S}) - 2 \epsilon} \right) \int_0^1 \tilde{y}_t^2 dt + 
% \Paren{\frac{\epsilon(4+3\mu(\rset))}{\mu(\rset)^2-\eps^2}}^2 \\
%	& \le \frac{1}{3} \cdot \left( \frac{\mu(\tilde{\rset})
%		+ \epsilon}{\mu(\tilde{\rset}) - 
%		2\epsilon}\right)^2\frac{(1 - \mu(\tilde{\rset}))^2}{2} \\
%	& \hspace{5mm}+  \Paren{\frac{\epsilon(4+3\mu(\rset))}{\mu(\rset)^2-\eps^2}}^2 \\
%	& \le \frac{1}{3} \cdot \left(\frac{\mu(\tilde{\rset})
%		+ \epsilon}{\mu(\tilde{\rset}) - 2\epsilon}\right)^2
%	\frac{(1 - \mu(\rset) + \epsilon)^2}{2} \\
%	& \hspace{5mm}+   \Paren{\frac{\epsilon(4+3\mu(\rset))}{\mu(\rset)^2-\eps^2}}^2,
%\end{align}
%\fi
   which, since
 \begin{equation}
     \int_0^1 \tilde{y}_t dt = 0.
 \end{equation}
 implies that
 \begin{align}
     \int_0^1 y_t^2 dt & \le \frac{1}{3}\left(\frac{1 - \mu(\tilde{\rset})}{2}
               \right)^2  + 4\epsilon^2 \\
         & \le \frac{1}{3}\left(\frac{1 - \mu(\rset) + \epsilon}{2}
               \right)^2 + 4\epsilon^2,
   \end{align}
 from \eqref{eq:ramp:close2}.
 Taking $\epsilon \rightarrow 0$ establishes \eqref{eq:rampineq}.

 Now given any code, $\rampenc$, for the ramp, 
 define $S_i$ as in \eqref{eq:ramp:Sdef}.
 Then we have from \eqref{eq:ramp:expansion}
 \begin{align}
     \dist\left( \rampenc \right) & = \sum_i \mu(\rset_i)
     \Paren{\frac{1}{12} - \int_0^1 \EE[J_t|\phase \in \rset_i]^2 dt} \\
     & \ge \sum_i \mu(\rset_i) \Paren{\frac{1}{12} - \frac{1}{3}
         \left(\frac{1 - \mu(\rset_i)}{2}\right)^2} \\
         & = \sum_i \mu(\rset_i) \Paren{\frac{1}{6}\mu(\rset_i) 
     - \frac{1}{12} \mu^2(\rset_i)} \label{eq:rampd_intervals}.
 \end{align}
 Setting $p_i = \mu(\rset_i)$ establishes the converse of the
 theorem. Achievability follows by noting that, for any
 $\{p_i\}$ satisfying the constraint in \eqref{eq:rampopt2}, we
 can create a code  with entropy $-\sum_i p_i \log p_i$
 satisfying the distortion constraint by setting $\rset_0,
 \rset_1, \ldots, \rset_{n-1}$ to be intervals with $\mu(\rset_i) = p_i$
 and noting equality in \eqref{eq:rampineq}.
\end{proof}
\else \begin{proof}[Proof Sketch] Each value of the phase variable defines a unique ramp realization which is mapped to a codeword by a given ramp encoder $\rampenc$. For $i \in \NN$, let $\rset_i \subset \left[0,1\right]$ be the quantization cell of $i$. The entropy can be written in terms of the Lebesgue measure of $\rset_i$'s. The main part of the proof involves bounding the distortion as
	\[ \dist\left( \rampenc \right) \geq \sum\limits_i \mu(\rset_i) \Paren{\frac{1}{12} - \frac{1}{3} \Paren{\frac{1-\mu(\rset_i)}{2}}^2 }  \]
for an arbitrary measurable $\rset_i$ with equality holding if each of the $\rset_i$'s are intervals. \end{proof}\fi

The entropy-distortion function in \eqref{thm:rampED} is formally
an infinite-dimensional optimization problem.
The following lower bound is easily computed.

\begin{theorem}
	\begin{equation}
		\eramp(D) \geq \sup\limits_{\lambda \geq 0} \inf_{0<p_i<1} -\log(p_i) + \lambda \frac{p_i(2-p_i)}{12}.
	\end{equation} 
    \label{thm:ramp:lower}
\end{theorem}
\ifnum0\ifdefined\EXTENDED 1\fi \ifdefined\JOURNAL 1\fi >0
\begin{proof}
	By weak duality,
	\begin{align*}
		\eramp(D) &\geq \sup\limits_{\lambda \geq 0} \inf_{\substack{\left\{ p_i \right\}_{i=1}^{\infty}: \\ \sum_i p_i = 1, \\ p_i \geq 0 \text{ for all } i}} \sum\limits_{i=1}^{\infty} -p_i\log(p_i)  +\lambda \frac{p_i^2(2-p_i)}{12} \\
		&= \sup\limits_{\lambda \geq 0} \inf_{\substack{\left\{ p_i \right\}_{i=1}^{\infty}: \\ \sum_i p_i = 1, \\ p_i \geq 0 \text{ for all } i}} \sum\limits_{i=1}^{\infty} p_i\Paren{-\log(p_i)  +\lambda \frac{p_i(2-p_i)}{12}} \\
		&\geq \sup\limits_{\lambda \geq 0} \inf_{0<p_i<1} -\log(p_i) + \lambda \frac{p_i(2-p_i)}{12}.
	\end{align*}
\end{proof}
\else \fi 

The bound in Theorem~\ref{thm:ramp:lower} is illustrated in Fig.~\ref{fig:r_RD} (``lower bound'').
\ifdefined\EXTENDED
 For rates equal to $\log K$ where $K\geq 2$ is an integer, an achievable scheme uniformly quantizes the phase. The encoder assigns a ramp realization $J_t$ to the midpoint of the interval that contains $J_0 + 0.5$. The decoder outputs the conditional mean of the ramp restricted to the phase lying in the encoded interval. For rates between $\log K$ and $\log K+1$, an upper bound on the optimal tradeoff is obtained by dividing the ramp into biuniform intervals, where one interval has length $\eps$ and the other $K$ intervals have length $\frac{1-\eps}{K}$.
 \else
 An upper bound can be obtained by quantizing $V$ using quantization cells with a biuniform size distribution
 (``achievable'' in Fig.~\ref{fig:r_RD}).
 \fi

We next compare this optimal performance
against that of stochastically-trained
ANN-based compressors.

\subsection{ANN performance}
%TODO: Revisit once sawbridge section is transferred. 
\ifdefined\JOURNAL
\begin{figure}
	\centering
	\includegraphics[width=\linewidth]{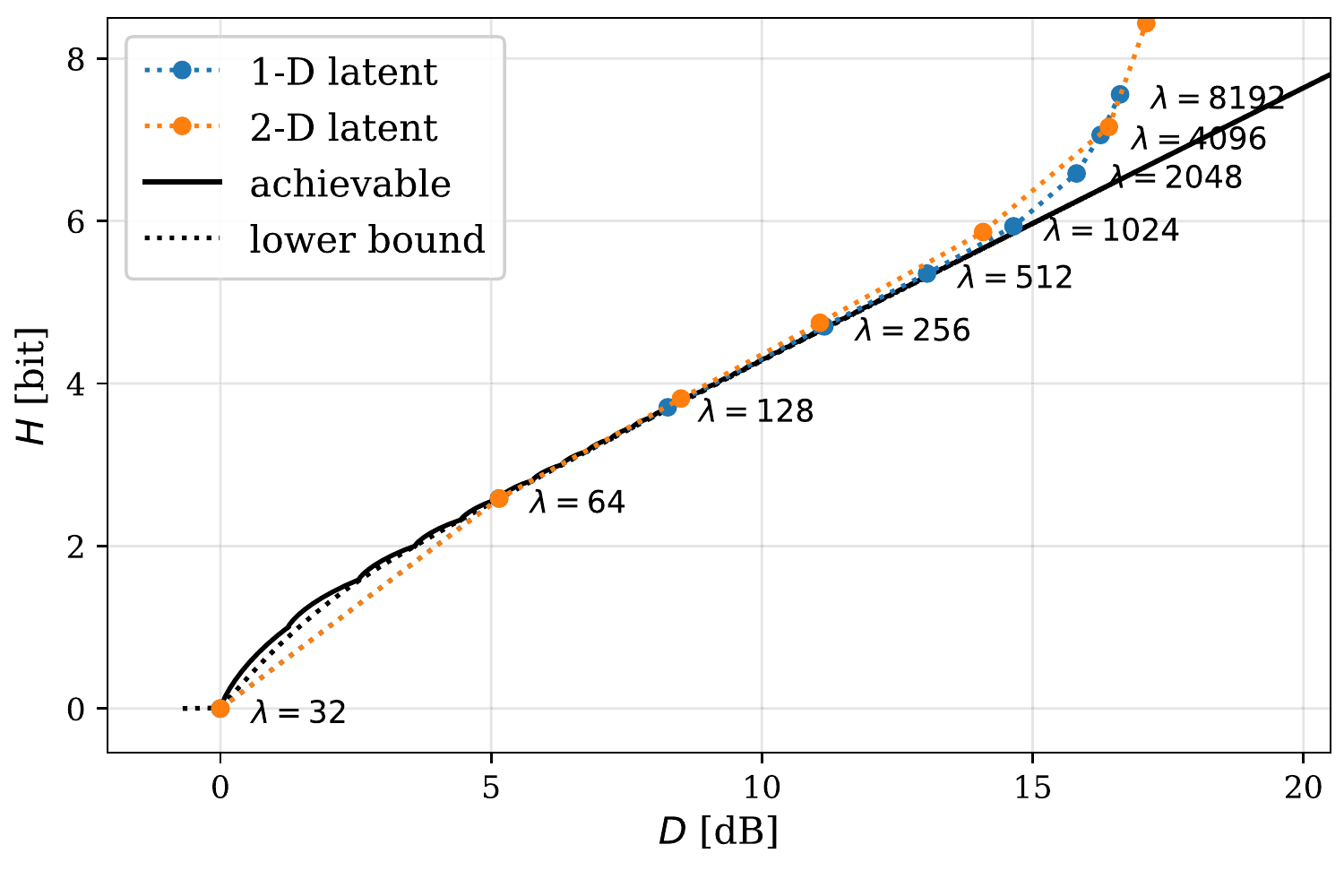}
	\caption{Entropy-distortion tradeoff for ramp.}\label{fig:r_RD}
\end{figure}
\else
\begin{figure}[!tb]
	\centering
	\includegraphics[width=\linewidth]{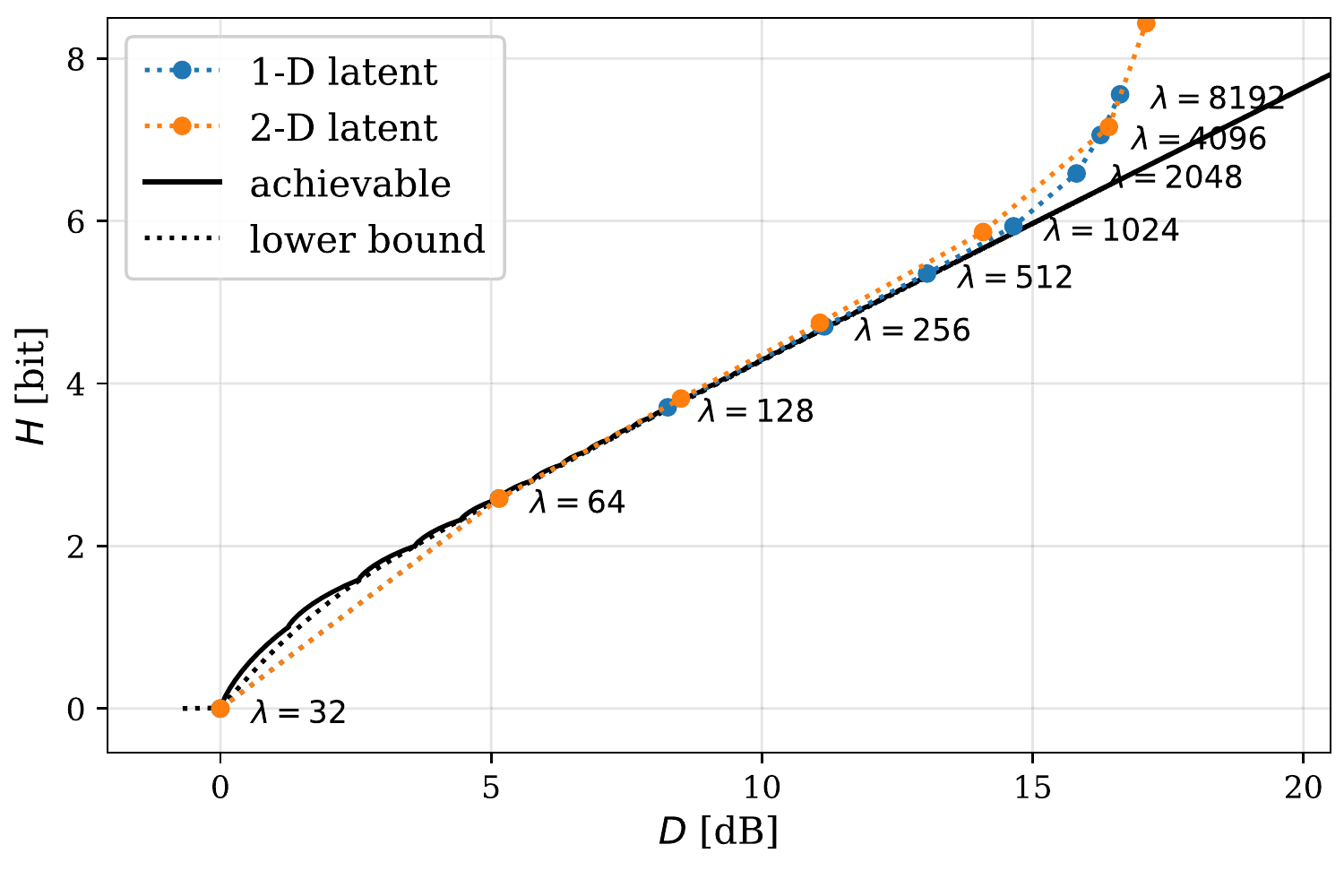}
	\caption{Entropy-distortion tradeoff for ramp.}\label{fig:r_RD}
\end{figure}\fi

%We see that the performance is suboptimal 
%at high rates. 
%Thus, unlike the 
%sawbridge~\cite{WaBa:Sawbridge:Long}, the ANN-based compressor does not optimally compress the ramp.
%\begin{figure*}[!tb]
%	\centering
%	\begin{minipage}{.5\textwidth}
%		\centering
%	%\includegraphics[scale=0.35]{tmp/r_anl}
%	\includegraphics[scale=0.40]{r_1D_latent_low_rate}
%	\end{minipage}%
%	\begin{minipage}{0.5\textwidth}
%	\centering
%	\includegraphics[scale=0.40]{r_1D_low_rate_op}
%	\end{minipage}
%	\caption{Ramp, $\lambda = 64$. Left: Analysis transform as a function of the phase. Right: Reconstruction at randomly chosen index as a function of phase. The orange plot is obtained by passing the output of the analysis transform through the synthesis transform without quantization.}\label{fig:r_low}
%\end{figure*}
%
%\begin{figure*}[!tb]\label{fig:ramp_high}
%	%TODO: Same sizes for figures.
%	\centering
%	\begin{minipage}{.45\textwidth}
%			\centering
%		%\includegraphics[scale=0.35]{tmp/r_anl}
%		\includegraphics[scale=0.40]{r_1D_latent_high_rate}
%	\end{minipage}
%	\begin{minipage}{0.45\textwidth}
%		\centering
%		\includegraphics[scale=0.4]{r_1D_high_rate_op}
%	\end{minipage}
%	\caption{Ramp, $\lambda=4096$. Left: Analysis transform as a function of the phase. Right: Reconstruction at randomly chosen index as a function of phase.}\label{fig:r_high}
%\end{figure*}

\begin{figure*}[!tb]
	\centering
%	\begin{subfigure}{.24\textwidth}
%		\centering
%		%\includegraphics[scale=0.35]{tmp/r_anl}
%		\includegraphics[scale=0.25]{r_1D_latent_low_rate}
%		\caption{}\label{fig:r_low_lt}
%	\end{subfigure}%
%	\begin{subfigure}{0.24\textwidth}
%		\centering
%		\includegraphics[scale=0.25]{r_1D_low_rate_op}
%			\caption{}\label{fig:r_low_op}
%	\end{subfigure}
	\begin{subfigure}{0.43\textwidth}
		\centering
		\includegraphics[scale=0.41]{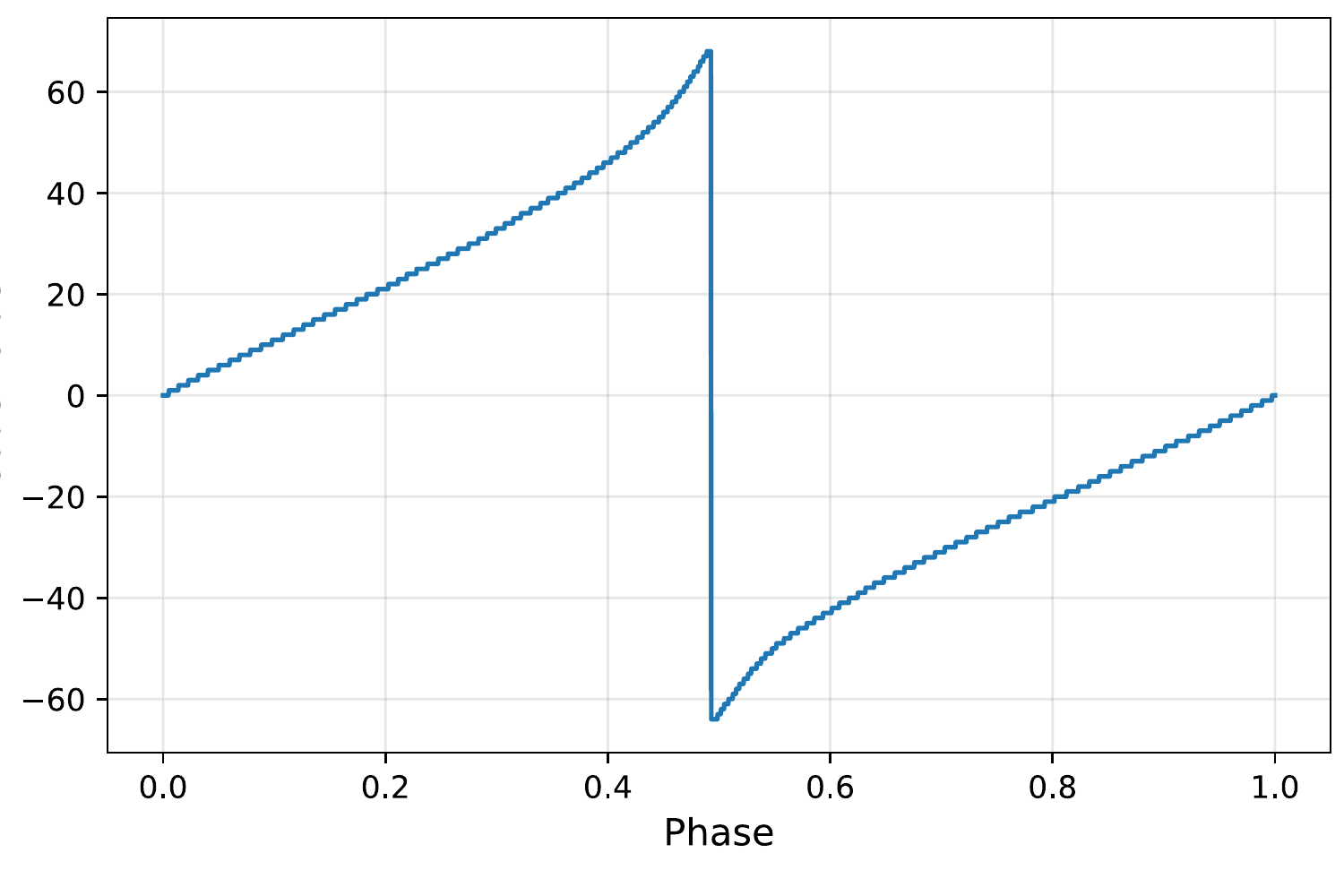}
			\caption{}\label{fig:r_high_lt}
	\end{subfigure}
	\begin{subfigure}{0.43\textwidth}
		\centering
		\includegraphics[scale=0.41]{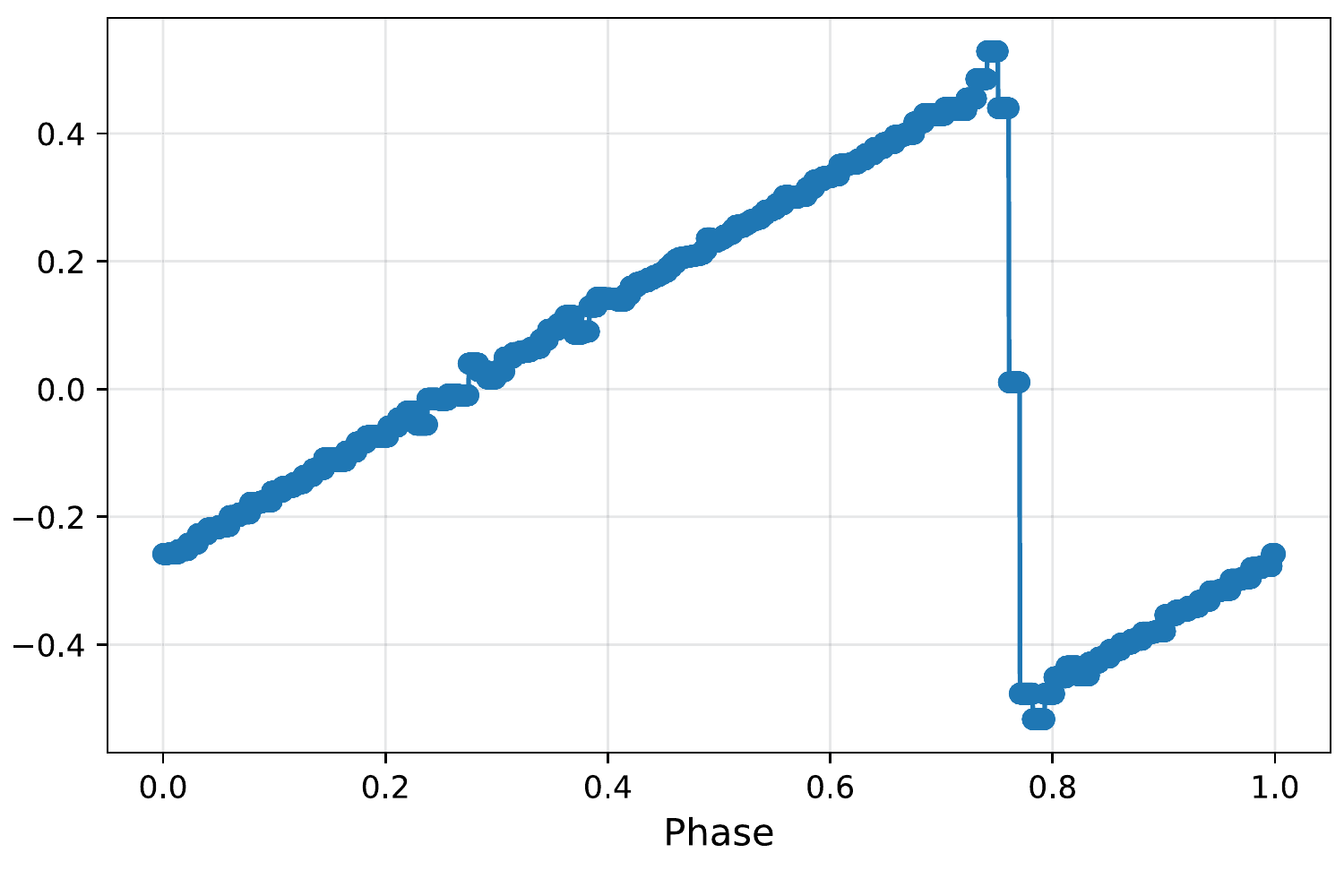}
			\caption{}\label{fig:r_high_op}
	\end{subfigure}
	\caption{%(a) Codeword value vs phase for $\lambda=48$ (on optimal tradeoff). (b) Reconstruction at randomly chosen index (240) vs phase for $\lambda=48$. The unquantized plot is obtained by passing the output of $\anl$ through $\syn$ without quantization. 
        (a) Quantized encoder output vs. phase for $\lambda=4096$ (away from optimal tradeoff). (b) Reconstruction at randomly chosen index vs. phase for $\lambda=4096$. Here the synthesis transform is insufficiently steep.}
\end{figure*}

Figure~\ref{fig:r_RD} shows the entropy-distortion tradeoff 
for the ANN-based compressor.
We see that the performance is suboptimal at high rates.
The problem is similar to that which arose with the circle.
Per Theorem~\ref{thm:rampED},
we expect the analysis transform to output the phase $\phase$,
or some scaled and shifted version of it such as 
$Y = \alpha(v + \phase \mod 1) + \beta$. The training process does indeed
find such a solution for $\anl(\cdot)$, as shown
in Fig.~\ref{fig:r_high_lt}. Now consider the synthesis transform,
$\syn(\cdot)$, evaluated at a particular output time,
$t$. The mapping we require in order to recover
$J_t$ from $Y$ is 
\begin{equation}
    \label{eq:ramp:desired}
    Y \mapsto ((Y-\beta)/\alpha + t - v) \mod 1.
\end{equation}
This function is not continuous, however, and being
a multilayer perceptron, $\syn(\cdot)$ must be 
continuous (and indeed differentiable).  In practice, 
the trained synthesis transform will implement a 
continuous approximation to (\ref{eq:ramp:desired}) that 
incurs distortion because its discontinuity is 
insufficiently steep. \ifdefined \JOURNAL Thus the Lipschitz-limiting condition affects the ramp too. However, unlike the circle, it manifests in the synthesis transform. \else The situation is thus analogous
to the circle. Note that for the circle, however, the
problem arose with the analysis transform, whereas
here it arises in the synthesis transform. \fi Indeed,
evaluating the ramp at one fixed time $t$ 
already has the desired form for the analysis
transform,
\begin{equation}
    \rmp_t = \left[ \left( t + \phase \right) \text{ mod } 1  \right] - \frac{1}{2}.
\end{equation}
Thus the analysis transform can simply transmit
any of the input samples directly,
which is clearly a continuous map.
On the other hand,
Fig.~\ref{fig:r_high_op}
shows the reconstruction, evaluated
at a fixed time index, as a function of 
$\phase$. We see that the discontinuity
is insufficiently steep.

As with the circle, at low rates, the approximation error noted 
above is negligible compared with the quantization error,
and thus this phenomenon does not lead to entropy-distortion
suboptimality.

\ifdefined\JOURNAL
This reasoning assumes that the synthesis transform must be 
of the form in (\ref{eq:ramp:desired}), which we have
not established. We have merely shown, via the proof of 
Theorem~\ref{thm:rampED}, that it is an optimal approach. In the 
appendix, we argue more generally that the analysis and 
synthesis transforms cannot be both optimal and continuous.
\fi

\section{Overparametrizing the Latent Space}
\label{sec:2D}

One possible workaround to the suboptimality identified
above is to overparametrize the latent space. In 
the above examples, the analysis transform could
output a 2-D vector ($\cdim = 2$) rather than a 
scalar, for instance. In the case of the circle,
this evidently solves the identifiability problem,
as the analysis transform can now simply be the
identity map. This makes the quantization problem
more difficult, however, because the latent variables
are always quantized independently of each other, 
but if the analysis transform is indeed the identity map then
the two latent variables are not statistically
independent (though they are uncorrelated). With
an identity analysis transform, it is possible to
independently quantize the two latent variables
so that the overall system is optimal if one
uses quantizers that are highly asymmetrical between
the two components. Specifically, one of the components
is quantized to a single bit to indicate which
hemisphere the source is in, and the other 
is quantized with high resolution. The resulting
system will be optimal so long as the quantized
values are independent and uniformly distributed
over their respective supports. Remarkably,
with enough training, the network is able
to find this solution, as shown in Figs.~\ref{fig:c_RD}
and \ref{fig:c_anl_2D}. Of course, this only works
when the number of reconstructions is even. 

For the ramp, on the other hand, overparametrizing
the latent space appears to provide little 
benefit (Fig.~\ref{fig:r_RD}).
From the discussion in Section~\ref{sec:ramp}, this is
unsurprising. 

\section{Discussion}
We conclude that the argument that ANN-based compressors
are adept at compressing low-dimensional manifolds should
be applied with some care. While ANNs are essentially
optimal compressors for some manifolds~\cite{WaBa:Sawbridge:Long}, they
are suboptimal for the sources exhibiting a circular
topology considered in this paper.

Low-level models of natural images have long included 
spatially-local image features that exist on one or more intrinsic dimensions and can be organized on circular topologies, such as spatial orientation
 \cite{SiFrAdHe92,OlFi96,BeSe97,HaSc98}.
It has also been noted that different edge or feature profiles can be determined by their phase, another dimension which has a circular topology, and their phase congruency, in complex wavelet representations~\cite{VeOw90,Ko99}.
Empirically, it has been shown that local image statistics are inherently
low-dimensional~\cite{HeBaRaSi15} and that these local statistics live on a
topology equivalent to a Klein bottle~\cite{CaIsSiZo08}, a topology which is
also found in orientation-selective complex wavelet filters.
Thus while the sources considered in this paper
are obviously synthetic, the findings in this paper 
may have some applicability to image compression.

Beyond the findings about ANN-based compressors, we have
also determined the entropy-distortion functions of the circle
and ramp. Classical rate-distortion theory has
favored the study of Gaussian sources over
those with low-dimensional manifold structure, despite
the relevance of the latter to the study of image compression.
Prior work on compression of low-dimensional sets includes
rate-distortion tradeoffs for fractal sources~\cite{KawabataDembo94} and rate-distortion analysis of compressed sensing \cite{WuVerdu10,LeinonenCJK18}
More connected to the results of this paper, \cite{RieglerKB18} provides a lower bound and \cite{KolianderPRH16} an upper bound on the rate-distortion function of the circle. We have provided a complete characterization of the entropy-distortion function.

%\ifdefined\EXTENDED
%\appendices
%\input{../journal/appendix}
%\fi

\printbibliography
%\bibliographystyle{IEEEtran}
%\bibliography{IEEEabrv,../journal/biblio}

\end{document}